\documentclass[useAMS,usenatbib,twocolumn,a4paper,fleqn]{mn2e}
\usepackage{amsmath,amssymb,cancel}
\usepackage{graphicx}
\usepackage{euscript}
\usepackage{bm,tabularx}
\usepackage[normalem]{ulem}  
\usepackage{float}
\usepackage[usenames,dvipsnames]{pstricks}
\usepackage{epsfig}
\usepackage{pst-grad} 
\usepackage{pst-plot} 
\usepackage{lipsum, mathtools}
\usepackage{comment}
\usepackage{enumerate}
\usepackage{wrapfig,multicol}
\usepackage{color,colortbl}
\usepackage[colorlinks,citecolor=blue,linkcolor=red,urlcolor=blue]{hyperref}
\newcommand{\EQ}{\begin{equation}}
\newcommand{\EN}{\end{equation}}
\newcommand{\EQA}{\begin{eqnarray}}
\newcommand{\ENA}{\end{eqnarray}}

\newcommand{\meanrho}{\overline{\rho}}
\newcommand{\lf}{\ell_{\rm f}}

\newcommand{\ee}{\bm{e}}

\newcommand{\kk}{\bm{{{k}}}}

\newcommand{\rr}{\bm{{{r}}}}
\newcommand{\dd}{\mathrm{d}}
\newcommand{\EE}{\bm{{{E}}}}
\newcommand{\FF}{\bm{{{F}}}}
\newcommand{\LL}{\bm{{{L}}}}
\newcommand{\BB}{\bm{{{B}}}}

\newcommand{\UU}{\bm{{{U}}}}
\newcommand{\JJ}{\bm{{{J}}}}

\newcommand{\meanrp}{\langle r_{p}\rangle}
\newcommand{\ted}{t_{\rm ed}}
\newcommand{\mrms}{\mathcal{M}_{\rm rms}}
\newcommand{\brms}{B_{\rm rms}}
\newcommand{\urms}{u_{\rm rms}}
\def\Rm{\rm Rm}
\def\Rmc{{\rm Rm}_{\rm cr}}
\def\Rey{{\rm Re}}
\def\Pm{\rm Pm}
\def\mrho{\bar{\rho}}
\def\mbmag{\bar{B}}

\newcommand{\mnras}{MNRAS}

\newcommand{\pre}{Phys.~Rev.~E}
\newcommand{\prl}{Phys.~Rev.~Lett.}
\newcommand\aap{Astron.\ Astrophys.}

\newcommand\apj{Astrophys.\ J.}
\newcommand\apjs{Astrophys.\ J.\ S.}
\newcommand\apjl{Astrophys.\ J.\ L.}

\newcommand\physrep{Phys.\ Rep.}
\newcommand\ssr{Space Sci. Rev.}

\title[Fluctuation dynamo saturation]{Role of magnetic pressure forces in fluctuation dynamo saturation}

\author[S.~Sur and K.~Subramanian]{
Sharanya Sur$^{1}$\thanks{E-mail: sharanya.sur@iiap.res.in}
and Kandaswamy Subramanian$^{2,3}$
\\
$^{1}$Indian Institute of Astrophysics, 2nd Block, 100 Feet Road, Koramangala, Bangalore 560034, INDIA\\
$^{2}$IUCAA, Post Bag 4, Ganeshkhind, Pune 411007, INDIA\\
$^{3}$Department of Physics, Ashoka University, Rajiv Gandhi Education City, Rai, Sonipat 131029, Haryana, INDIA
}
\date{Accepted 2023 November 13. Received 2023 November 13; in original form 2023 May 16}

\pubyear{\the\year{}}

\begin{document}
\label{firstpage}
\pagerange{\pageref{firstpage}--\pageref{lastpage}}
\maketitle

\begin{abstract}

Using magnetohydrodynamic simulations of fluctuation dynamos in 
turbulent flows with rms Mach numbers $\mrms = 0.2, 1.1$ 
and $3$, we show that magnetic pressure forces play a crucial role in 
dynamo saturation in supersonic flows. First, as expected when 
pressure forces oppose compression, an increase in anticorrelation 
between density and magnetic field strengths obtains even in subsonic 
flows with the anti-correlation arising from the intense but rarer magnetic 
structures. In supersonic flows, due to stronger compressive motions 
density and magnetic field strength continue to maintain a positive 
correlation. However, the degree of positive correlation decreases as the 
dynamo saturates. Secondly, we find that the unit vectors of $\nabla\rho$ 
and $\nabla B^{2}$ are preferentially antiparallel to each other in subsonic 
flows. This is indicative of magnetic pressure opposing compression. This 
antiparallel alignment persists in transonic and supersonic flows at dynamo 
saturation. However, compressive motions also lead to the emergence of 
a parallel alignment in these flows. Finally, we consider the work done 
against the components of the Lorentz force and the different sources of 
magnetic energy growth and dissipation. We show that while in subsonic 
flows, suppression of field line stretching is dominant in saturating the 
dynamo, the picture is different in supersonic flows. Both field line 
stretching and compression initially amplifies the field. However,
growing magnetic pressure opposes further compression of magnetic flux 
which tends to reduce the compressive motions. Simultaneously, field 
line stretching also reduces. But, suppression of compressive 
amplification dominates the saturation of the dynamo.

\end{abstract}



\section{Introduction}
\label{intro}

It is now well established that dynamically insignificant seed magnetic fields 
embedded in a conducting fluid can be amplified to near equipartition strength 
by a three-dimensional random or turbulent flow 
\citep{B50,K68,HBD04,Scheko+04a,BS05,SSH06,Fed+11,BS13,SPS14,
PJR15,Fed16,XL16,SBS18,R19,Seta+20,XL20,SF21b,T21,SS21}.
The growth of the field depends on competition between inductive growth
and resistive dissipation which translates to the requirement that the 
magnetic Reynolds number ($\Rm$) is above a critical instability threshold 
$\Rmc$. This process, referred to as fluctuation dynamo (FD) amplifies 
seed fields exponentially fast (on eddy-turnover time-scales) by stretching 
and compression of field lines driven by random (in time) motions that 
is supplied by a random/turbulent flow, until a saturation process becomes 
important. FDs are crucial to explain the possible presence of magnetic 
fields in elliptical galaxies \citep{MS96,Sur19,Seta+21}, in young galaxies 
at high redshifts \citep{SBS18} and in clusters of galaxies 
\citep{SSH06, CR09, BS13, PJR15, Marinacci+18, Vazza+18, Donnert+18, 
Sur19, SS21,SBS21}. In addition, they are also likely to operate in disc galaxies, 
generated from turbulent fluid motions \citep{Pak+17, Gent+23} and provide 
seed fields for large-scale dynamo action. 

Given the ubiquity of the occurrence of FDs in astrophysical objects, we revisit 
the mechanism of FD saturation in turbulent flows in this work. The standard 
picture of the non-linear evolution of the dynamo is the following. For magnetic 
Prandtl numbers $\Pm > 1$ and fluid Reynolds number ($\Rey$) large enough 
such that the velocity spectrum consists of multiple scales, the dynamo is excited 
initially by random motions on small enough scales $\ell<< l_{\rm f}$, where 
$\Rm(\ell) > \Rmc$ 
\citep{HBD04, Scheko+04a, SS21}, where $l_{\rm f}$ is the driving scale of 
turbulence. The amplification of the field due to such eddies continue till the 
local magnetic energy density approaches values comparable to the energy 
density of fluid motions on similar scales. Further growth of the field by such 
eddies is suppressed due to non-linear back reaction arising from the Lorentz 
force. However, turbulent eddies at a neighbouring larger scale can still continue 
to grow the field as they have a larger energy. Gradually over time, scale-by-scale 
saturation of the field occurs with the peak of the magnetic energy spectrum 
approaching a fraction of the outer scale (or the forcing scale of turbulent 
motions) \citep[][and references therein]{SS21}. 

Although the above mechanism has been confirmed in numerical simulations 
by a number of authors mentioned earlier, the exact manner in which the 
approach to a global non-linear steady state occurs still remains unclear. For 
this purpose it is instructive to consider a qualitative picture of FD amplification. 
In an ideal plasma, $\BB/\rho \propto \delta\rr$, where $\BB$ is the magnetic 
field, $\rho$ the plasma density and $\delta\rr$ the infinitesimal separation 
vector of fluid elements along the field line \citep{BS05}. If $\delta\rr$ increases 
in a mean square sense, due to random stretching of the field, and if $\rho$ 
is assumed strictly constant as would obtain in an incompressible flow, the 
magnetic energy also increases. Suppression of random stretching of the 
magnetic field lines by the growing magnetic tension (of the Lorentz force) 
has been recognised as the key mechanism ushering in the saturation of the 
dynamo \citep{Sch+02}. 

However, there is another possibility to limit growth in case we do not impose 
strict incompressibility, which obtains due to the effect of magnetic pressure. 
When a region containing a flux of field lines (or a flux tube) of length '$L$' is 
stretched, magnetic pressure opposes the compression of the flux tube 
cross-section of area '$A$'. Consequently, $\rho$ decreases to conserve 
mass, and the magnetic field strength $B$ then need not increase. This 
mechanism could be important for FD saturation in the interstellar medium 
(ISM) of galaxies, where turbulence is expected to be supersonic at the 
driving scale. Analysis of the alignment between the magnetic field and the 
eigenvectors of the rate of strain tensor ($S_{ij}$) in compressible magnetohydrodynamic 
(MHD) simulations by \cite{SPS14b} further showed that compressive motions 
are statistically oriented perpendicular to the direction of stretching. In a similar 
vein, magnetic pressure can also oppose any purely compressive motions 
and prevent the increase of $\rho$ hence $B$. We examine here how 
important is this mode of saturation for the FD in random compressible flows, 
even if the compressibility effects are small as in subsonic flows. Note, 
however, that the role of magnetic pressure in saturating the dynamo 
cannot be captured in numerical simulations which solves the exactly 
incompressible MHD equations. 
 
The paper is organised as follows. In Section~\ref{methods}, we present
in brief the initial conditions and the set-up of the simulations. Thereafter, 
results of our investigation is presented under various sections. To guide 
the readers, we begin by first discussing the evolution of the rms Mach 
number of the flows, the ratio of magnetic to kinetic energies, and the PDFs 
of the magnetic energy density in Section~\ref{emag}. With the overarching 
goal to elucidate the role of 
magnetic pressure in FD saturation, we then proceed to explore the time 
evolution of the $\rho-B$ correlation and how regions with varying field 
strengths (as obtained in the FD) and over densities contribute to this 
correlation in Section~\ref{correlation}. If magnetic pressure plays a role 
as described earlier, we expect this correlation to decrease as the field 
becomes dynamically important. Next, in Section~\ref{align}, we study in 
detail the nature of the alignment between the magnetic field direction, 
the gradient of the density, and the components of the Lorentz force, 
namely the gradient of the magnetic pressure and the magnetic tension 
and explore how these alignments manifest themselves in the kinematic 
and non-linear stages of dynamo evolution. In Section~\ref{impact}, we 
study the work done against the magnetic tension and gradient of the 
magnetic pressure and how it underlines the importance of magnetic 
pressure in dynamo saturation. Next, similar to previous studies \cite[e.g.,][]{Seta+20,SF21b}, 
we also explore the relative importance of stretching, advection, compression 
and dissipation on the growth and saturation of magnetic energy in 
Section~\ref{local_effects}. Finally, in Section~\ref{conclu} we conclude 
with highlights of the key results and emphasise the role of magnetic 
pressure forces in dynamo saturation.

\section{Numerical method and initial conditions}
\label{methods}

We focus our analysis on three non-ideal MHD simulations of non-helically 
forced FD simulations in the simplest possible numerical 
setup. The simulations were performed with an isothermal equation of state 
and periodic boundaries using the {\small{FLASH}} code\footnote{https://flash.rochester.edu/site/flashcode/} 
\citep{Fry+00,Benzi+08, EP88} (version 4.2), which solves the three-dimensional 
compressible MHD equations using explicit viscosity and resistivity 
in a conservative form. However, to better elucidate the role of the 
different terms, we express the MHD equations in a 
non-conservative form here : 
\EQA 
\partial_{t}\rho + \nabla\cdot(\rho\UU) &=& 0, \nonumber \\ 
\rho\left[\partial_{t}\UU + (\UU\cdot\nabla)\UU\right] &=& (\BB\cdot\nabla)\BB - \nabla p' + \nabla\cdot(2\nu\rho\mathcal{S}) + \rho\FF, \nonumber \\
\partial_{t}\BB + (\UU\cdot\nabla)\BB &=& (\BB\cdot\nabla)\UU - \BB(\nabla\cdot\UU) + \eta\nabla^{2}\BB, \\ \nonumber 
\nabla\cdot\BB &=& 0. 
\ENA 

Here, $\rho$, $\UU$, $p'=p+ \left|\BB\right|^2/2$, $\BB$, denote density, 
velocity, total pressure (thermal and magnetic) and the magnetic field, respectively. 
$\mathcal{S}_{ij}=(1/2)(\partial_i u_j+\partial_j u_i)-(1/3)\delta_{ij}\partial_{k}u_{k}$ 
is the traceless part of the rate of strain tensor, $\rho\FF$ is the artificial driving 
term and $\nu$ and $\eta$ are the constant viscosity and resistivity, respectively.

In all the simulations, turbulence is driven solenoidally (i.e., $\nabla\cdot\FF = 0$) 
over a range of wave numbers between $1 \leq |\kk|L/2\pi \leq 3$ (such that the 
average forcing wave number $k_{\rm f}L/2\pi = 2$ and '$L$' is the length of the box) 
as a stochastic Orstein-Ulhenbeck (OU) process with a finite time correlation. To 
cover different regimes of compressibility, we analyse the data from a subsonic, 
a transonic and one supersonic simulation with steady state rms Mach 
numbers $\mrms = u_{\rm rms}/c_{\rm s} = 0.2, 1.1$ and $3.0$, 
respectively.  Here $u_{\rm rms}$ is the turbulent rms velocity and $c_{s}$ is the
isothermal sound speed. All simulations have a resolution of $512^{3}, \Pm = \Rm/\Rey = 1$
and are in dimensionless units. Both $\Rey$ and $\Rm$ are defined 
with respect to the driving scale of turbulence. The initial conditions 
consist of a box of unit length, the density $\rho$ and $c_{\rm s}$ set to unity 
with zero initial velocities. The magnetic field is 
initialised as $\BB_{\rm init} = B_{0}[0,0,\sin(10\pi x)]$, with $B_{0}$ adjusted 
to a value such that the initial plasma beta $\beta_{0} = p_{\rm th}/p_{\rm mag} \approx 10^{6} - 10^{7}$. 
Thus, the magnetic field is dynamically weak to start with in all the 
simulations. Further, to maintain $\nabla\cdot\BB$ to machine precision 
level, we use the unsplit staggered mesh algorithm in {\scriptsize FLASHv4.2} 
with a constrained transport scheme \citep{LD09, Lee13} and a Harten-Lax-van 
Leer discontinuities (HLLD) Riemann solver \citep{MK05}. Table~\ref{tab:sumsim} 
lists the important parameters of these runs. 

\begin{table}
\centering 
\setlength{\tabcolsep}{10.0 pt}
\caption{Key parameters of simulations used in this study. The resolution 
in each case is $512^{3}$. $k_{\rm f}L/2\pi = 2 $ is the average forcing 
wave number and $\mrms$ is the average values of the rms Mach 
number in the steady state. $\ell_{\rm f} = 2\pi/k_{\rm f}$ is the forcing scale and 
$\Pm$ and $\Rey$ are the magnetic Prandtl number and the fluid Reynolds 
numbers, respectively.}
\begin{tabular}{ccccc} \hline
Run & $k_{\rm f}L/2\pi$ & $\mrms$ & $\Pm$ & $\Rey = u\,\lf/\nu$ \\ \hline
A & 2.0 & $0.2$ & 1.0 & 1500 \\  
B & 2.0 & $1.1$ & 1.0 & 1250 \\  
C & 2.0 & $3.0$ & 1.0 & 2250 \\ \hline
\end{tabular}
\label{tab:sumsim}
\end{table}

\section{Magnetic energy evolution and its probability distribution 
function} 
\label{emag}

\begin{figure}
\centering
\includegraphics[width=\columnwidth]{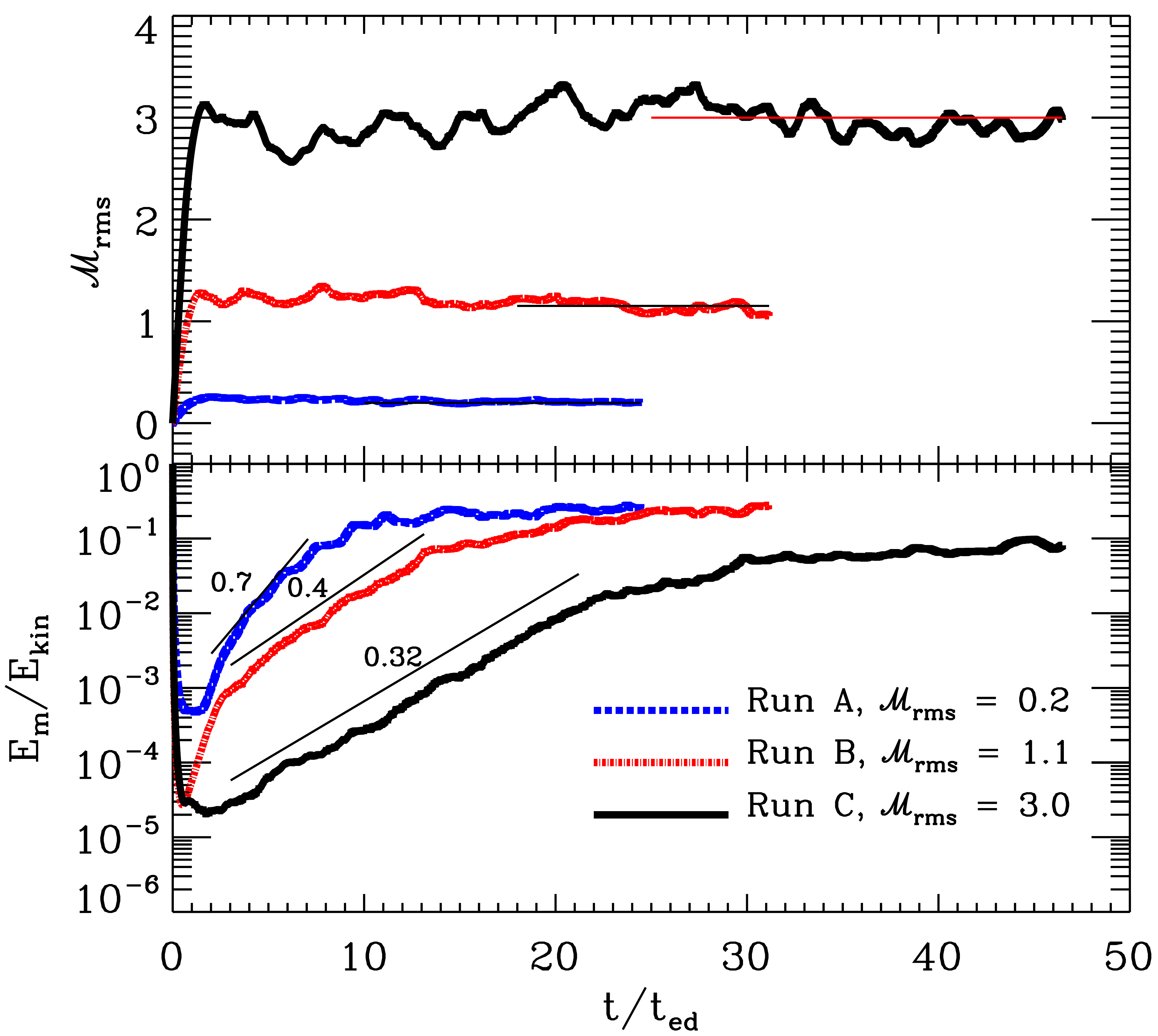} 
\caption{Time evolution of $\mrms$ (top panel) and the ratio of $E_{\rm m}/E_{\rm k}$ 
(bottom panel) for all the runs. The thin horizontal lines in the top panel are the average 
values of $\mrms$ in the time interval considered. The thin black lines in the bottom 
panel denote the slopes of $E_{\rm m}/E_{\rm k}$ in the kinematic phase}. 
\label{fig:emag_evol}
\end{figure}

Before delving into the details on the role of magnetic pressure forces in 
dynamo saturation, it is worthwhile to explore some of the important evolutionary 
features that we find in our simulations. Specifically, we show in Fig.~\ref{fig:emag_evol} 
the evolution of the rms Mach number (top panel) and the ratio of magnetic to 
the kinetic energies (bottom panel) as a function of the eddy-turnover time 
($\ted = l_{\rm f}/u_{\rm rms}$). It takes about $(1-2)\ted$ before a steady 
state of turbulence is established. The thin horizontal lines in the top panel 
are the average values of $\mrms$ which are estimated in the intervals 
$t/\ted = (10-25), (18-31)$ and $t/\ted = (25-47)$ for runs A, B and C, respectively. 
These values are also shown in Table~\ref{tab:sumsim}.

The bottom panel shows that the evolution of $E_{\rm m}/E_{\rm k}$ has three 
distinct phases : an initial exponential phase (kinematic) followed by an intermediate 
stage of slower growth and eventual saturation. It is clearly evident that the 
kinematic phase lasts for about $t/\ted = 8, 13$ and $22$ in runs A, B and C, 
respectively. The slope of the curves in this phase are denoted by thin solid 
lines. Thereafter, an intermediate phase of slower growth ensues. The onset 
of saturation in runs A, B and C occurs from $t/\ted = 15, 22$ and $30$ respectively. 
This can also be inferred from the evolution of the magnetic spectra $M(k)$ in 
Fig.~\ref{fig:spec_comp} where the spectra tend to bunch together from the aforesaid 
times. The bottom panel further shows that steady state value of $E_{\rm m}/E_{\rm k}$ 
decreases as the compressibility of the flow increases. In the interval $t/\ted = 15 - 25$, 
the steady state value of $E_{\rm m}/E_{\rm k} = 0.23$ for run A, while for run B 
and C the steady state values are $0.22$ (for $t/\ted = 22 - 31$) and $0.07$ 
(for $t/\ted = 30 - 47$), respectively.

\begin{figure}
\centering
\includegraphics[width=\columnwidth]{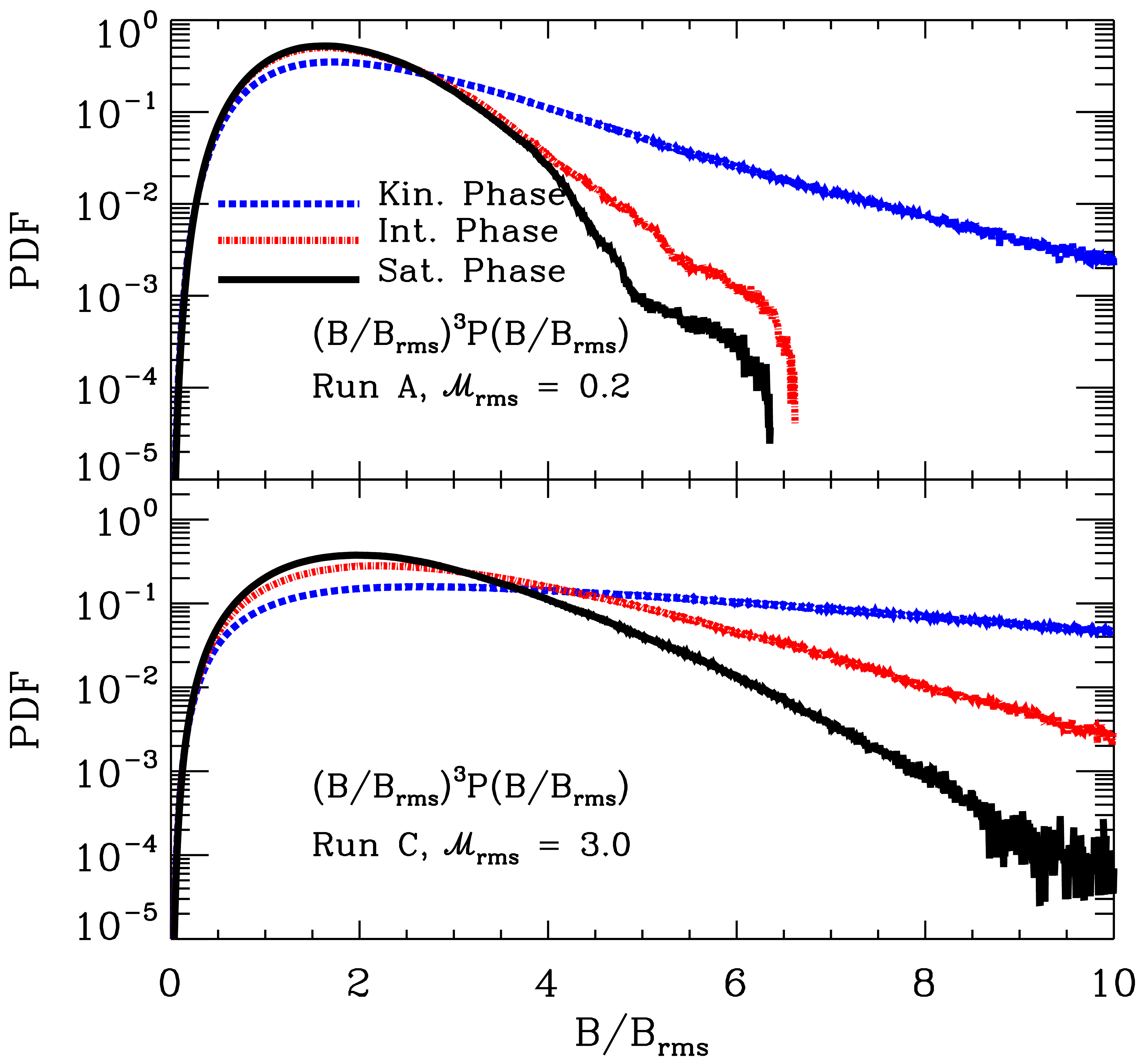} 
\caption{Total PDFs of $(B/\brms)^{3}P(B/\brms)$ in the kinematic (blue, 
dashed), intermediate (red, dash-dot) and saturated 
phases (black, solid) for runs A (top panel) and C (bottom panel).}
\label{fig:pdfs_bmag}
\end{figure}

A distinct feature of FD is that it generates both less intense volume filling 
fields together with much rarer but strong fields. To illustrate our point, 
we show in Fig.~\ref{fig:pdfs_bmag}, the probability distribution functions (PDFs)
of $(B/\brms)^{3}P(B/\brms)$ in the kinematic (blue, dotted lines), intermediate
(red, dash-dotted) and saturated phases (black, solid lines) for run A (subsonic) 
in the top panel and run C (supersonic) in the bottom panel. Note that 
$(B/\brms)^{2}$ represents the normalized magnetic energy and multiplying it 
by an additional factor of $B/\brms$ to obtain $(B/\brms)^{3}$ measures the 
magnetic energy density contributed by fields in a logarithmic interval of 
$B/\brms$. Here $P(B/\brms)$ is the PDF of the magnetic field strength. 
Two features are evident from these plots. First, in the kinematic phase, 
magnetic fields of a wide range of strengths are present and contribute to 
the total energy. Especially in the supersonic case, we find fields with strengths 
$\approx 10\brms$ contribute comparable energy density to the total as 
the rms fields. This is not the case with the subsonic run. This feature 
could be due to the continuous amplification of fields by density compression 
in addition to amplification by stretching. It is further evident that in both 
intermediate and saturated phases, the peak of the PDFs in both subsonic 
and supersonic cases lies at $\approx 2\,B/\brms$. However, the 
probability of fields $B > 6\brms$ contributing to the total energy decreases 
gradually from the kinematic to intermediate and saturated phases in the 
supersonic case. Strong field regions with $B/\brms > 2$ also contribute 
to the magnetic energy with a probability of $10^{-2}$ for $B/\brms = 6$ in 
the supersonic case. This implies that together with the general sea of volume 
filling fields, strong field regions may also play a role in dynamo saturation, 
particularly when compressibility effects are significant. We also note that 
in the supersonic case, regions with much higher field strengths ($B/\brms > 6$) 
continue to exist albeit with a much lower probability. This could imply that 
even though the dynamo suppresses the compressive motions when the 
fields become dynamically important, density compression still manages 
to amplify fields in the rarer regions.

\section{Correlation between density and magnetic field strength} 
\label{correlation}

In our simulations, density and magnetic field strength are uncorrelated to 
start with. To explore the temporal evolution of the (anti)correlation between
them as the FD evolves, we compute the Pearson correlation coefficient 
($r_{p}$) defined as, 
\begin{align} 
r_{p}(\rho, B) = \frac{{\rm Cov}(\rho, B)}{\sigma_{\rho}\sigma_{B}} &  \\ & \hspace{-2cm}\nonumber
= \frac{\sum_{i,j,k}(\rho_{i,j,k} - \mrho)(B_{i,j,k} - \mbmag)}
{\sqrt{\sum_{i,j,k}(\rho_{i,j,k} - \mrho)^{2}}\,\sqrt{\sum_{i,j,k}(B_{i,j,k} - \mbmag)^{2}}},
\label{rp}
\end{align}
where $B = \sqrt{B_x^{2} + B_{y}^{2} + B_{z}^2}$ is the magnitude of 
the field, $\rho_{i,j,k}$ and $B_{i,j,k}$ are the density and the magnetic 
field strength at a point $(i,j,k)$ in the simulation volume. $\mrho$ 
and $\mbmag$ are the mean values of density and $B$, respectively. 

\begin{figure}
\includegraphics[width=\columnwidth,height=6cm]{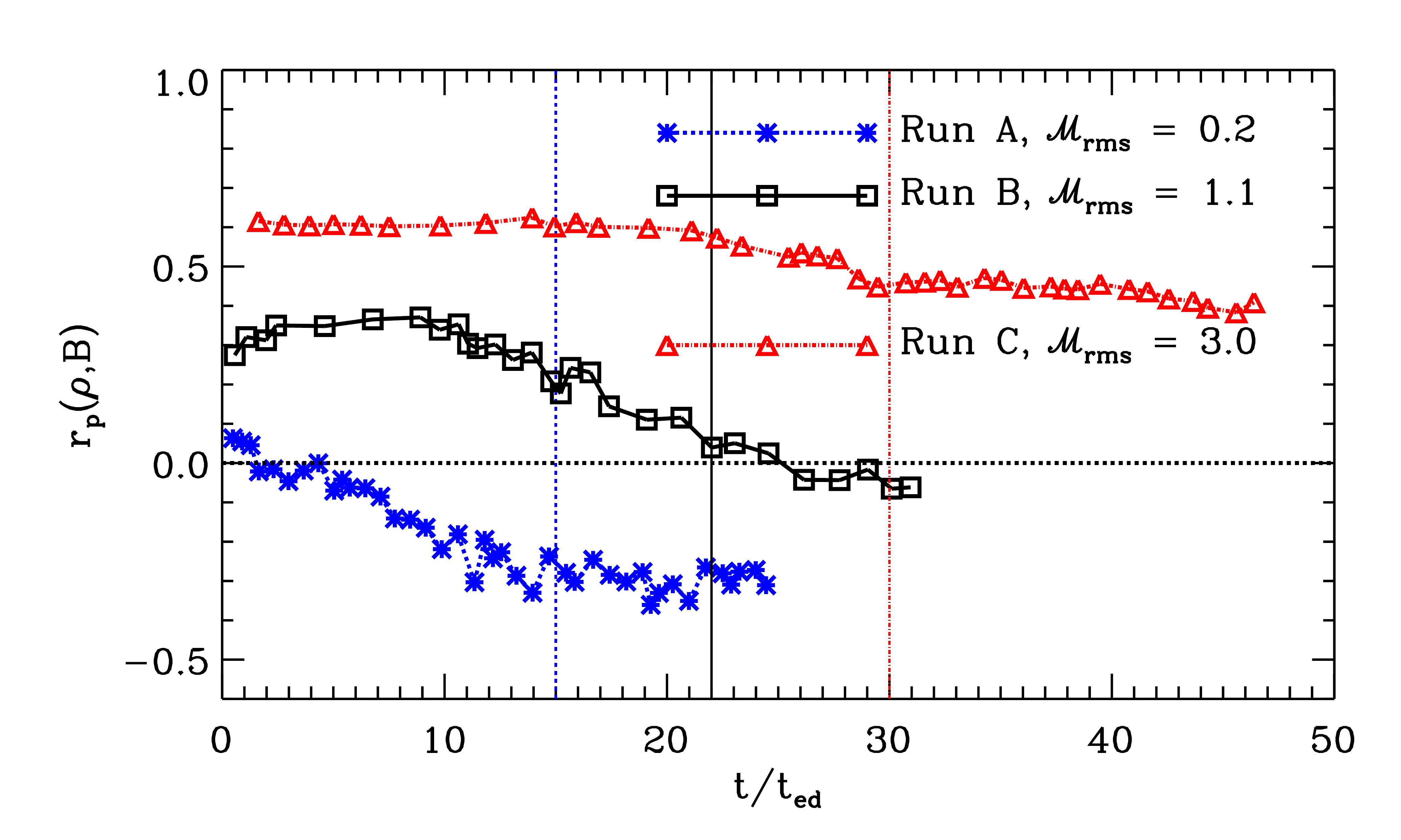}
\caption{Evolution of $r_{p}(\rho,B)$ for runs A, B and C. The blue dotted, black solid,
and red dashed vertical lines indicate the time of transition to the saturation phase 
in the respective runs.}
\label{fig:corr_sub_trans_super}
\vspace{-1.0em}
\end{figure}

Fig.~\ref{fig:corr_sub_trans_super} shows that in the subsonic case (run A, 
blue dotted line with asterisks) $\rho$ and $B$ evolves 
to become anticorrelated after about one eddy-turnover time and eventually 
settles at an average value $\meanrp = -0.3$ in the saturated state of dynamo. 
Thus, when the effects of compressibility are weak, high-density regions 
correspond to low magnetic field strength regions and vice versa. This is a 
consequence of the magnetic pressure acting to oppose the compression 
of the flux tube. 

When compressibility effects become important (with the increase in $\mrms$) 
the growth of the field is expected to be driven by a combination of amplification 
by random stretching and compression in converging flows which also 
enhances the density. The former amplifies fields in regions where the 
vorticity is strong and the latter in regions where over-densities are large. 
This is because, the solenoidal nature of turbulent driving leads to 
the production of vortical motions that drive field amplification by random 
stretching \citep{HBD04, Fed+11,PJR15, CFTB21} in contrast to compressive 
motions. However, regions of strong vorticity may not correspond to regions 
with large over-densities. In the kinematic phase, when fields are dynamically 
weak, amplification due to density compressions seem to dominate the 
evolution of the correlation co-efficient. This leads to an initial positive 
correlation with the degree of correlation strongly dependent on the rms 
$\mathcal{M}$ number of the flow as shown also by \cite{YCK16} 
and \cite{SF21a}. 

Indeed, in the transonic case (run B, black line with squares), we observe 
a positive correlation with $\meanrp = 0.33$ which then gradually 
declines as the dynamo saturates. After about $25\ted$, the density and 
the magnetic field strength becomes uncorrelated with $\meanrp = -0.03$ 
in the saturated phase. Thus, with exception to the initial positive correlation, 
the evolution of $r_{p}(\rho, B)$ in the transonic case bears resemblance to 
the corresponding evolution in the subsonic case. The decrease in $r_{p}$ 
can be understood as a consequence of magnetic pressure forces gaining 
in importance, which acts to further resist the compressive motions resulting 
in suppression of density enhancements. This results in regions of strong 
fields amplified by random stretching, that are not necessarily associated with 
high density regions which arose from compression.

On the other hand, in the supersonic case (run C, red dotted lines with triangles) 
where $\delta\rho/\mrho \propto \mathcal{M}^{2} \sim 9$, the dominance of 
density compressions amplifying the field pushes the $\meanrp = 0.6$ 
up to $\approx 20\,\ted$, which corresponds to the duration of the kinematic 
phase. Thereafter, the correlation decreases only slightly as the field evolves 
to the non-linear saturated state with $\meanrp = 0.43$ in the interval 
$t/\ted= (30 - 47)$. This suggests that when compressible effects are strong, 
field amplification due to density compression dominates the evolution of 
$r_{p}$ even when Lorentz forces are strong enough to be dynamically 
significant.

\subsection{Evolution of the correlation coefficient for different ranges in 
$B/\brms$ and over densities}
\label{diffB}

A generic feature of FD is that it generates magnetic fields consisting of 
rarer, intense field structures embedded in a sea of less intense, volume 
filling fields. Therefore, it is worthwhile to distinguish the contribution 
to $r_{p}(\rho,B)$ seen in Fig.~\ref{fig:corr_sub_trans_super} from regions 
with differing magnetic field strengths. Note here that for a given range 
of $B/\brms$ we focus only on those regions in the simulation volume 
where the range condition is satisfied. Thus, the data in each of the ranges 
discussed below correspond to a subset of the full data set and 
equation (2) is solved for the respective data sets. 

\begin{figure}
\includegraphics[width=\columnwidth]{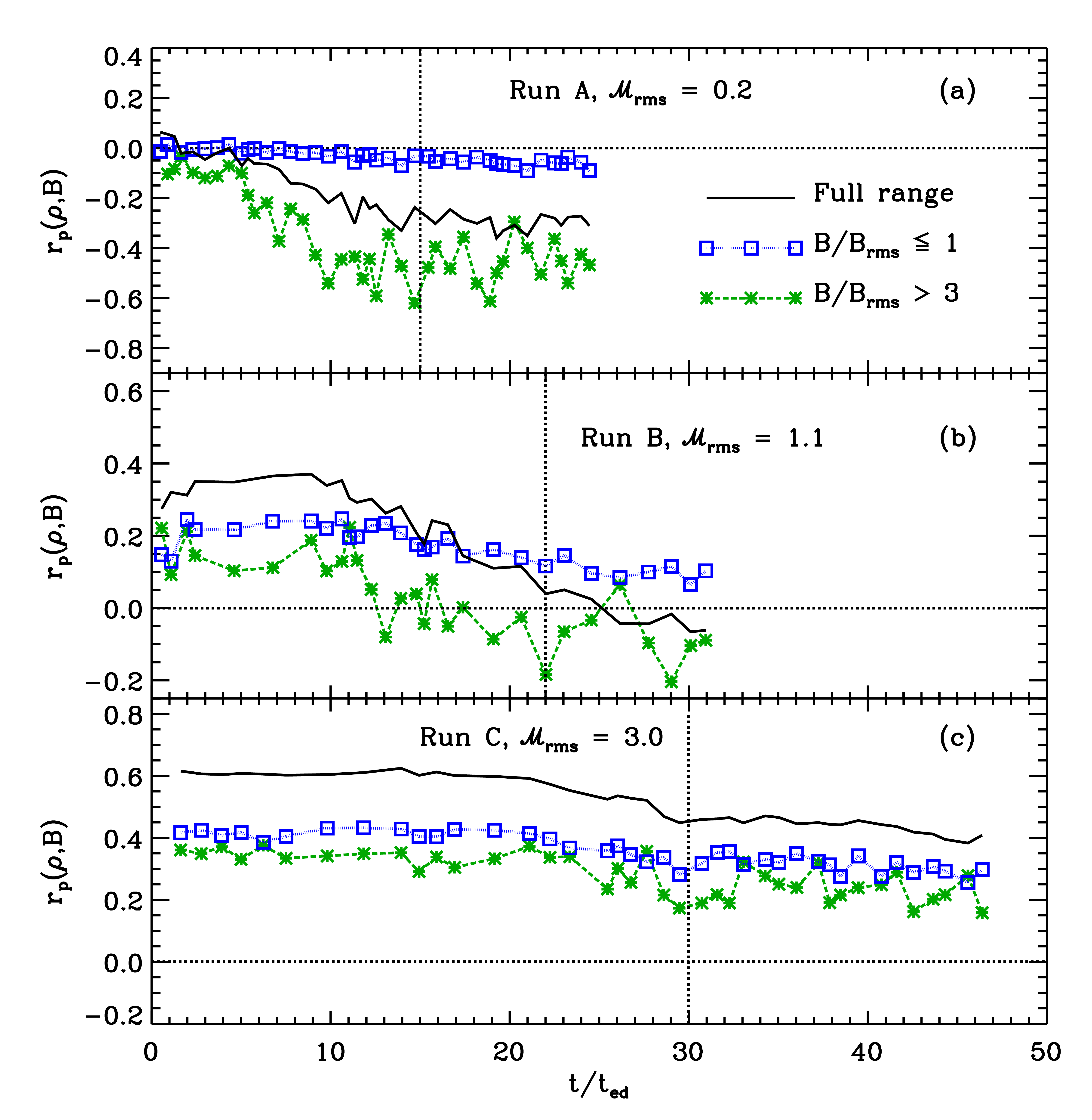}
\caption{Evolution of $r_{p}(\rho,B)$ for runs A, B and C for different ranges 
of $B/\brms$. The dotted vertical lines denote the transition to the saturated 
phase in the respective runs.}
\label{fig:bcuts}
\vspace{1.0em}
\end{figure}

In Fig.~\ref{fig:bcuts}, we show the evolution of $r_{p}$ for the subsonic 
(top panel), transonic (middle) and supersonic (bottom panel) for two different 
ranges of $B/\brms$ while the black solid lines in each panel depicts the 
variations of $r_{p}$ as in Fig.~\ref{fig:corr_sub_trans_super}, where no such 
ranges in $B/\brms$ are considered. 
For the subsonic case, it is now clearly evident that the strong anticorrelation 
$\meanrp = -0.46$ in the range $t/\ted = 13 - 25$ arises from the rarer, intense 
field regions where $B/\brms > 3$ (green dashed line with asterisk). In the 
absence of strong density compressions, such intense magnetic field regions 
solely arise due to random stretching. Thus, the observed strong anticorrelation 
implies that magnetic pressure exerted by these strong fields resist further 
compression of flux tubes. At the same time, the sea of less intense volume 
filling fields, however, remain uncorrelated with the density at all times. 

In the transonic case, initial density enhancements due to compression also 
enhances the magnetic field. This leads to an initial positive correlation, 
although the degree of positive correlation is lesser when regions with fields 
$B/\brms > 3$ are considered. However, the plots clearly show that once 
the dynamo builds up the fields, the degree of positive correlation steadily 
decreases even in regions where $B/\brms \leq 1$, while for the strong field 
regions, $r_{p}(\rho, B)$ evolves to be anticorrelated. Thus, with exception 
to the initial positive correlation, the general trend of $r_{p}(\rho,B)$ is very 
similar to the one observed in the subsonic case. Finally, in the supersonic 
case $\rho$ and $B$ remain positively correlated throughout the evolution. 
However, regions with $B/\brms > 3$ only show a weaker degree of positive 
correlation compared to regions with $B/B_{\rm rms} \leq 1$. 
This again reflects the fact that such strong and intense field 
regions arise due to amplification by field line stretching that obtains in 
vortical turbulence. However, density enhancements due to strong 
compressions continue to dominate over the magnetic field enhancements 
due to random stretching leading to a net albeit weaker positive correlation. 

\begin{figure}
\includegraphics[width=\columnwidth]{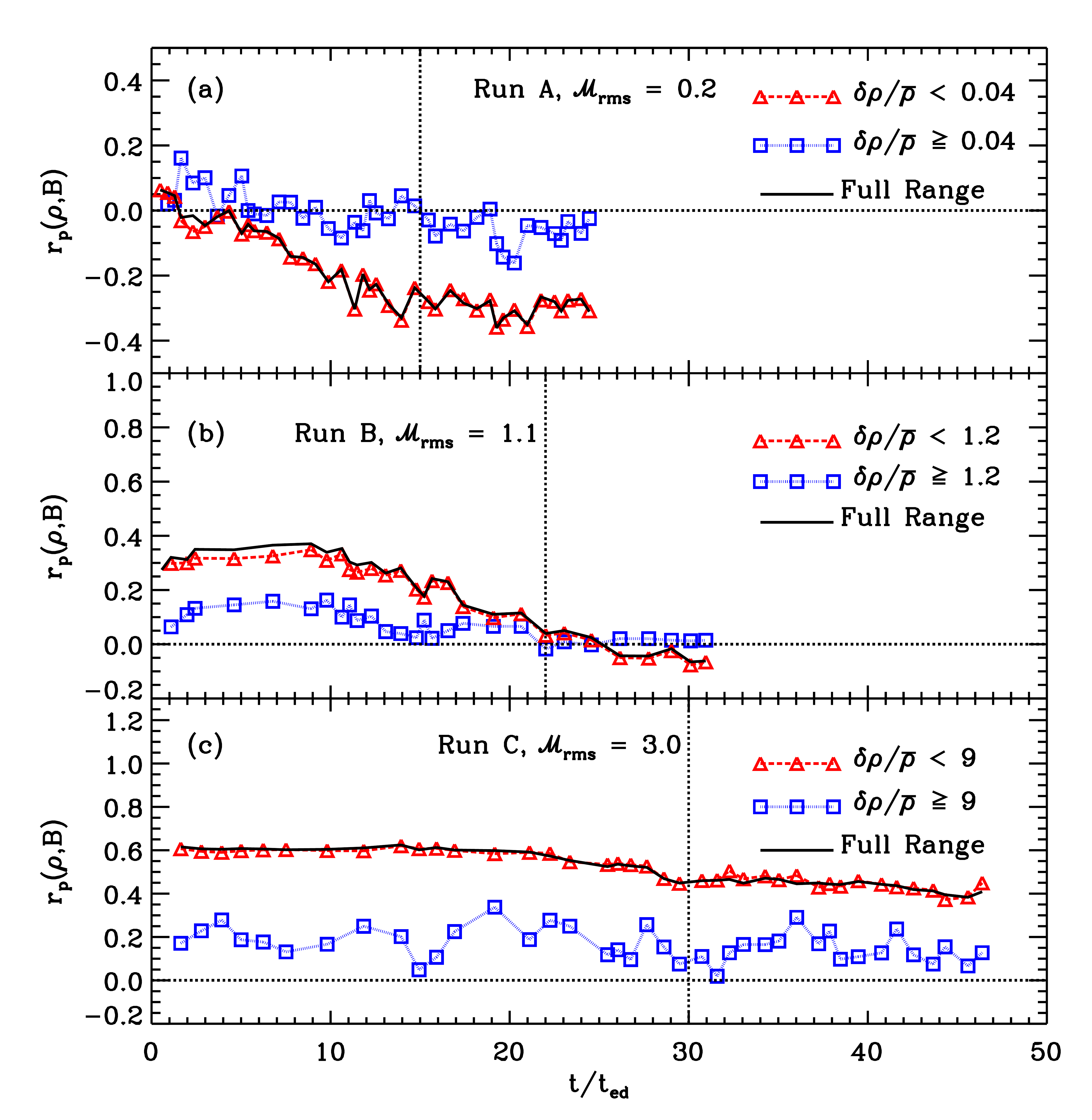}
\caption{Evolution of $r_{p}(\rho,B)$ for runs A, B and C for over density 
ranges segregated by regions : $\delta\rho/\meanrho < \mathcal{M}^{2}$ 
and $\delta\rho/\meanrho \geq \mathcal{M}^{2}$. As before, the dotted 
vertical lines in all the panels indicate the transition to the saturated phase 
in the respective runs.}
\label{fig:dcuts}
\vspace{1.0em}
\end{figure}

The fact that we consider three simulations with varying degrees of flow 
compressibility also makes it equally important to probe the degree of 
(anti)correlation in different over density ranges. Recall that density 
inhomegenities $\delta\rho/\meanrho \propto \mathcal{M}^{2}$, 
where $\delta\rho = (\rho - \mrho)$ is the fluctuation in density. While 
$\delta\rho/\meanrho$ is negligible in the subsonic case, significant density 
fluctuations $\propto \mathcal{M}^{2}$ could influence the correlation in 
transonic and  supersonic regimes. To explore this, we compute the evolution 
of $r_{p}(\rho, B)$ in density ranges segregated by regions : 
$\delta\rho/\meanrho < \mathcal{M}^{2}$ and 
$\delta\rho/\meanrho \geq \mathcal{M}^{2}$. 
The results are shown in Fig.~\ref{fig:dcuts} with panel (a) showing the 
evolution for the subsonic, panel (b) for transonic, and panel (c) for the 
supersonic run. 

We find that in the subsonic case, the observed anticorrelation arises 
from regions with $\delta\rho/\meanrho < 0.04$ as the magnetic fields in 
these regions are amplified solely due to vortical motions. Regions with 
$\delta\rho/\meanrho \geq 0.04$ remain uncorrelated. In the transonic 
and supersonic cases, the evolution of $r_{p}$ in regions with 
$\delta\rho/\meanrho < \mathcal{M}^{2}$ follow the pattern seen when no 
cutoffs are introduced (black solid lines). On the other hand, in both cases, 
$r_{p}$ starts out with a much weaker positive correlation when over dense 
regions with $\delta\rho/\meanrho \geq \mathcal{M}^{2}$ are considered. 
In the transonic case, magnetic field strength eventually becomes 
uncorrelated with the density while in the supersonic case, they maintain 
a weaker positive correlation. This could be due to the fact that such 
high-density regions are fewer in number in the simulation domain. 
Furthermore, we find that the values of $r_{p}$ calculated using a 
mass-weighted version of equation (2) are almost identical to the ones 
shown in Fig.~\ref{fig:dcuts} particularly for over densities 
$\delta\rho/\mrho > \mathcal{M}^{2}$.

\subsection{2D slices of density fluctuations and fluctuations in magnetic energy} 
\label{2dfluc} 

\begin{figure*}
\centering
\includegraphics[width=0.9\textwidth]{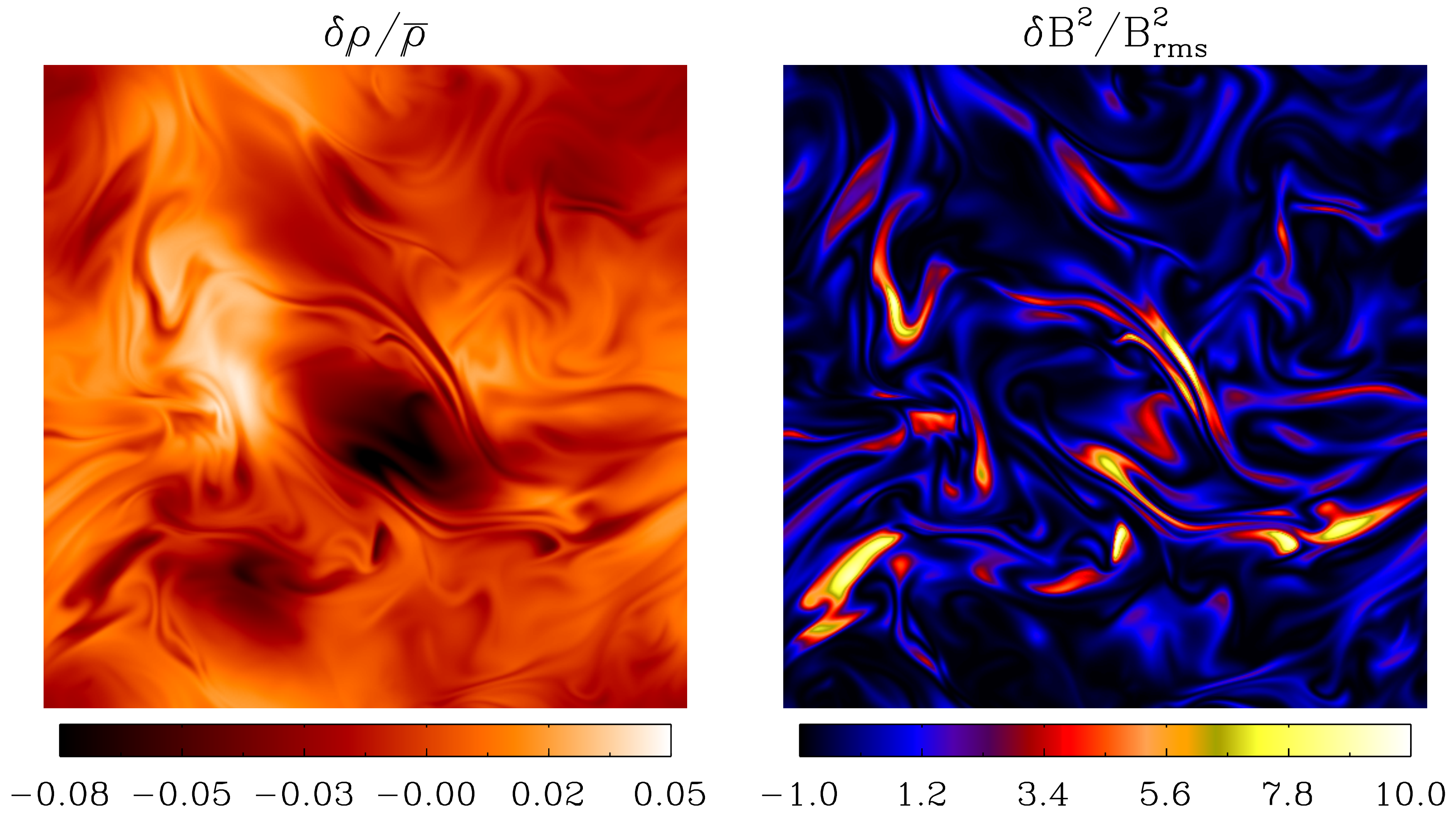}
\includegraphics[width=0.9\textwidth]{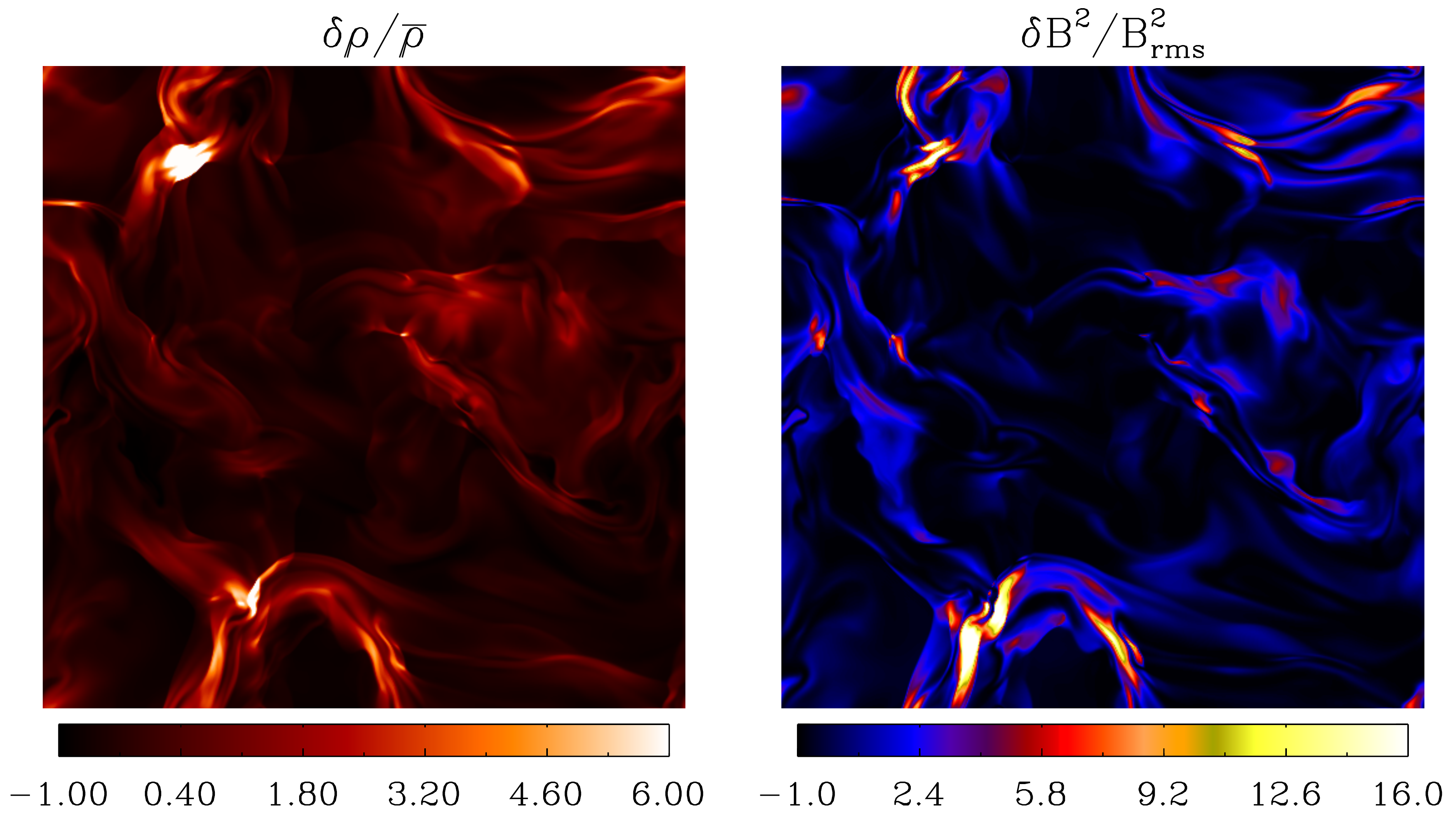}
\caption{2D slices of $\delta\rho/\mrho$ (left column) and $\delta B^{2}/B^{2}_{\rm rms}$ 
(right column) from snapshots in the saturated state at $t/\ted = 24.5$ from run A (top row)
and at $t/\ted = 48$ from run C (bottom row). The slices are shown in the $x-y$ plane 
at $z = 0$. The anticorrelation between the two in subsonic flows in clearly evident with 
bright regions of $\delta\rho/\mrho$ corresponding to dark regions (weaker fluctuations) 
in $\delta B^{2}/B^{2}_{\rm rms}$ and vice versa. On the other hand, the supersonic 
run exhibits positive correlation between $\delta\rho/\mrho$ and 
$\delta B^{2}/B^{2}_{\rm rms}$.
}
\label{fig:2d_slices}
\end{figure*}

Further evidence of the anticorrelation between the density and magnetic field 
strength obtained in run A can be easily discerned from the two-dimensional 
(2D) slices of the density and magnetic energy. Such slices can also reveal 
the positive correlation expected in supersonic flows. To illustrate this, we show 
in Fig.~\ref{fig:2d_slices}, 2D slices of $\delta\rho/\mrho$ (left column) and the 
corresponding fluctuations in $B^{2}$, 
$\delta B^{2}/B^{2}_{\rm rms} = (B^{2} - B^{2}_{\rm rms})/B^{2}_{\rm rms}$ 
(right column). These slices were obtained in the saturated phase of the 
dynamo in the respective runs. The top row shows the slices in the $x-y$ plane 
(at $z=0$) corresponding to $t/\ted = 24.5$ for run A, while the bottom row shows 
the same at $t/\ted = 48$ from run C.

It appears that fluctuations in both $\rho$ and $B^{2}$ resemble that of a 
cloth which is wrinkled and folded at multiple regions due to action of the 
turbulent driving. Note that even though both $\rho$ and $B^{2}$ are positive 
definite quantities, there are regions within the simulation volume where 
$\rho < \mrho$ ($\mrho = 1$) and $B^{2} < B^{2}_{\rm rms}$. Both the left 
and right panels of the figure shows a wide variety of high contrast but smoothly 
varying structures in $\delta\rho/\mrho$ and $\delta B^{2}/B^{2}_{\rm rms}$. 
More importantly, in run A we find that regions of strong density fluctuations 
correspond to regions of weak fluctuations in magnetic energy and vice versa. 
The strongest anti-correlation occurs in the rarer, strong field regions with 
$\delta B^{2}/B^{2}_{\rm rms} > 7.8$. This once again corroborates the 
results obtained in Section~\ref{diffB}. On the other hand, in comparison 
to run A the bottom row shows that fluctuations in both $\delta\rho/\mrho$ and 
$\delta B^{2}/B^{2}_{\rm rms}$ are much stronger in run C. It is also clearly 
evident that density fluctuations are positively correlated with fluctuations in 
magnetic energy in agreement with the results presented earlier.

\section{Alignment angles} 
\label{align}

\begin{figure*}
\centering
\includegraphics[width=\textwidth]{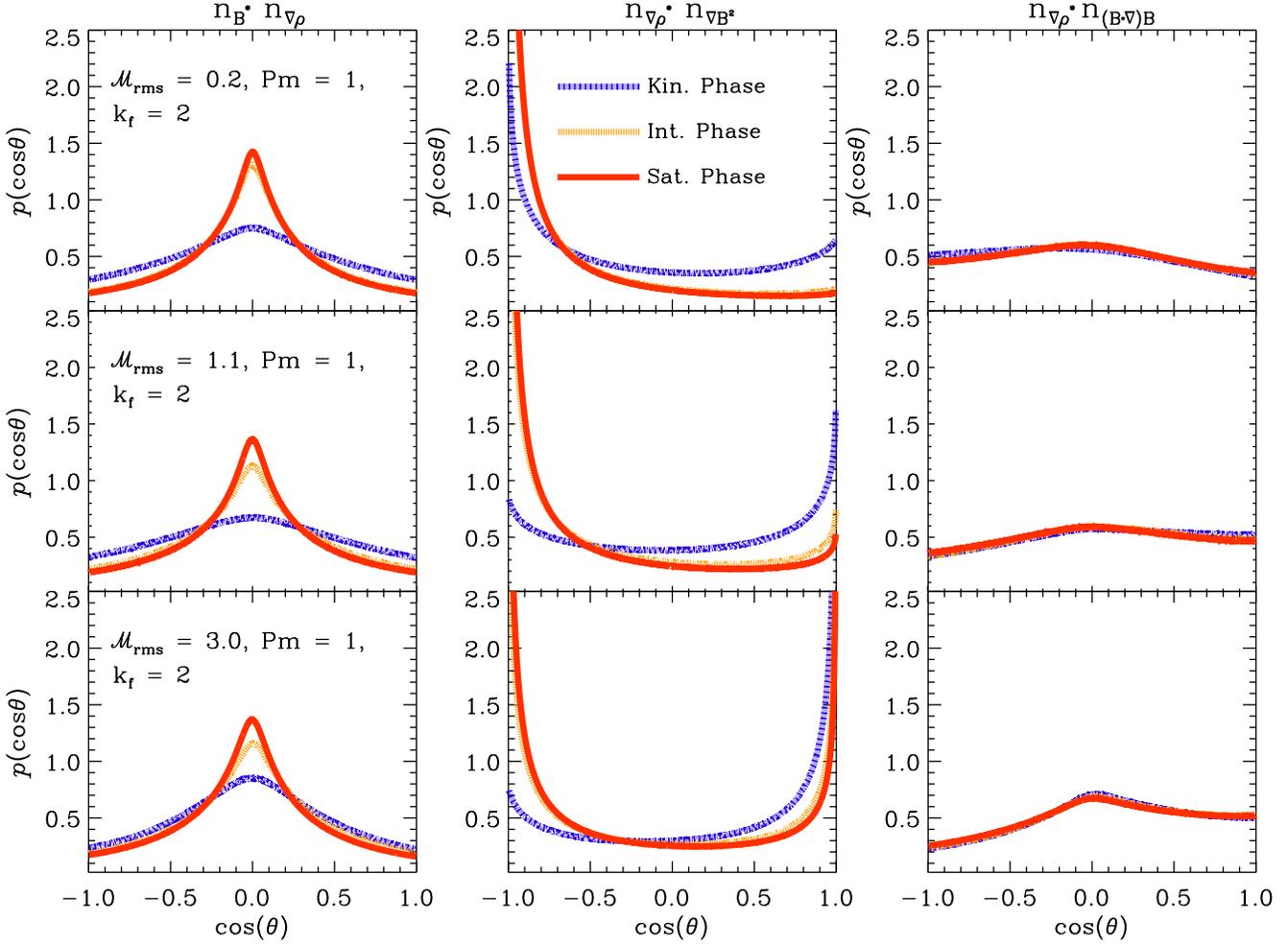}
\caption{PDFs of the cosine of the angles for the subsonic ($\mrms = 0.2$)
in the top row, transonic run ($\mrms = 1.1$) in the middle, and for 
the supersonic ($\mrms = 3.0$) run in the bottom row. The alignments 
are shown between : $n_{\BB}\cdot n_{\nabla\rho}$ (left column), 
$n_{\nabla\rho}\cdot n_{\nabla B^{2}}$ (middle column), and 
$n_{\nabla\rho}\cdot n_{(\BB\cdot\nabla)\BB}$ (right column).}
\label{fig:pdf_sub_super}
\end{figure*}

The evolution of $r_{p}(\rho,B)$ offers a first glimpse into the possible 
role of magnetic pressure in saturating the dynamo. Additionally, it is 
also worthwhile to investigate the nature of the alignments between the 
density gradient, magnetic field, and the components of the Lorentz force. 
Accordingly, we show in Fig.~\ref{fig:pdf_sub_super}, the PDFs of the 
cosine of the angles between the magnetic field and the gradient of 
density (left column), the gradient of density and gradient of the magnetic 
pressure (middle column), and finally, between the gradient of the density 
and the magnetic tension (right column). These PDFs are computed 
by averaging over a number of independent realisations (at different $\ted$)
in kinematic (blue, dash-dotted), intermediate (yellow, dashed) and saturated 
phases (red, solid). 

The left column of the above figure shows that the distributions of the 
unit vectors of $\nabla\rho$ and $B$ are symmetric about $\cos(\theta) = 0$ 
with the unit vector of the magnetic field (directed along the field line) 
preferentially perpendicular ($90^{\circ}$ or $270^{\circ}$) to the unit 
vector of the density gradient. This is consistent with the simple picture 
of a flux tube threaded by a field of strength '$B$' where the density can 
be enhanced or depleted in the flux tube. Moreover, the degree of 
orthogonal alignment appears to be stronger in the saturated and 
intermediate phases in comparison to the kinematic phase. 
This reinforces in a statistical sense, the simple qualitative 
picture described earlier. Due to magnetic pressure gradience opposing 
compression of the field, density is decreased in strong field regions 
and density contrasts are enhanced perpendicular to the field.
On the other hand, the gradient of the magnetic pressure ($\nabla B^{2}$) 
is directed inwards to the flux tube. The middle column in the figure shows 
that in the subsonic case (run A), $\nabla\rho$ and $\nabla B^{2}$ are 
antiparallel (i.e. at $180^{\circ}$) to each other in all the three phases.
This is a manifestation of the fact that as magnetic pressure gains in importance 
due to dynamo amplification, the force due to magnetic pressure opposes 
the compression of the flux tubes. This results in density variations that are 
anticorrelated with variations in magnetic pressure. Thus, force due to 
magnetic pressure tends to empty out the flux tubes rather than allowing the 
cross-sectional area to decrease. Note that in the subsonic case, this 
antiparallel alignment is caused by the rarer, stronger field regions that are 
strongly anticorrelated with the density, while the general sea of volume filling 
fields remain uncorrelated (see Fig.~\ref{fig:bcuts}). The kinematic phase in 
the transonic case (run B) shows an antiparallel alignment together with a 
weak parallel alignment between the two, which evolves to a stronger 
anti-parallel alignment in the intermediate phase and continuing to the 
saturated phase.

However, the picture changes considerably in the supersonic case (run C) 
where compressibility effects are strong enough that in the kinematic 
phase, density enhancements drive the amplification of the magnetic field. 
In this case, the strong positive correlation seen in Fig.~\ref{fig:corr_sub_trans_super} 
results in density variations that are also correlated with magnetic pressure 
variations. Consequently, $\nabla\rho$ and $\nabla B^{2}$ are initially 
aligned parallel to each other. However, over time random stretching
also amplifies magnetic fields in addition to field amplification due to 
compressions. Similar to the subsonic case, we find that as the dynamo 
saturates force due to magnetic pressure opposes further compression 
of the field lines resulting in the emergence of an antiparallel alignment 
between the two that coexists with the parallel alignment.  

The right column shows that magnetic tension does not show any 
preferred alignment with the density gradient in the subsonic case. 
However, with the increase in flow compressibility, a weak orthogonal 
alignment between the two emerges which becomes more prominent 
at $\mrms = 3$. Thus, the nature of the alignments discussed above 
provides further appreciation of the role of magnetic pressure in 
dynamo saturation.

\section{Impact of Lorentz forces on magnetic energy evolution}
\label{impact}  

In this section, we discuss how the energy exchange between the velocity 
and magnetic fields facilitated by the different components of the Lorentz 
force elucidates the role of magnetic pressure and magnetic tension in 
saturating the dynamo. 

The volume averaged magnetic energy evolution in terms of the work 
done by the Lorentz force is
\EQA
\label{tevol_emag1} 
\frac{d}{dt}\left\langle{|\BB|^{2}/2}\right\rangle = 
&-& \left\langle\UU\cdot(\JJ\times\BB)\right\rangle -\left\langle\frac{\JJ^{2}}{\sigma}\right\rangle \nonumber \\
&-& \left\langle\nabla\cdot(\EE\times\BB)\right\rangle
\ENA
where $\sigma$ is the conductivity and the angular brackets $\langle ...\rangle$ 
denotes volume averaging. Substituting 
$\JJ\times\BB = (\BB\cdot\nabla)\BB - \nabla(B^{2}/2)$ in the above equation 
and multiplying both sides by $\ted/\brms^{2}$, the magnetic energy evolution 
in dimensionless form is, 
\EQA
\label{tevol_emag2} 
\frac{d}{dt}\left\langle\frac{|\BB|^{2}}{2}\right\rangle\frac{t_{\rm ed}}{B_{\rm rms}^{2}} = 
&-&\left\langle\UU\cdot \LL_{\rm T}\right\rangle\frac{t_{\rm ed}}{B^{2}_{\rm rms}} \nonumber \\
&+&\left\langle\UU\cdot \LL_{\rm p}\right\rangle\frac{t_{\rm ed}}{B^{2}_{\rm rms}} 
-\left\langle\frac{\JJ^{2}}{\sigma}\right\rangle\frac{t_{\rm ed}}{B_{\rm rms}^{2}}. 
\ENA
The first two terms on the right-hand side correspond to the rate of work done by the 
magnetic tension $\LL_{\rm T} = \tilde{\LL}_{\rm T} - (\hat{\BB}\cdot\tilde{\LL}_{\rm T})\hat{\BB}$ 
and the gradient of the magnetic pressure $\LL_{\rm p} = \tilde{\LL}_{\rm p} - (\hat{\BB}\cdot\tilde{\LL}_{\rm p})\hat{\BB}$. 
Here, $\tilde{\LL}_{\rm T} = (\BB\cdot\nabla)\BB$ and $\tilde{\LL}_{\rm p} = \nabla(B^{2}/2)$ 
and $\hat{\BB}$ is the unit vector of the magnetic field. Thus, $\LL_{\rm T}$ 
and $\LL_{\rm p}$ as defined above only retain components perpendicular 
to $\BB$.
This is because the Lorentz force has no component parallel to $\BB$.
The last term represents the decrease in magnetic energy due to Joule 
dissipation.\footnote{The term proportional to the Poynting vector 
$\EE\times\BB$ goes to zero as the surface integral vanishes over the 
volume.} The effect of this term can be gauged from Figs~\ref{fig:pdf_mean_sub} 
and \ref{fig:pdf_mean_super} presented in the following section.

\begin{figure}
\centering
\includegraphics[width=\columnwidth]{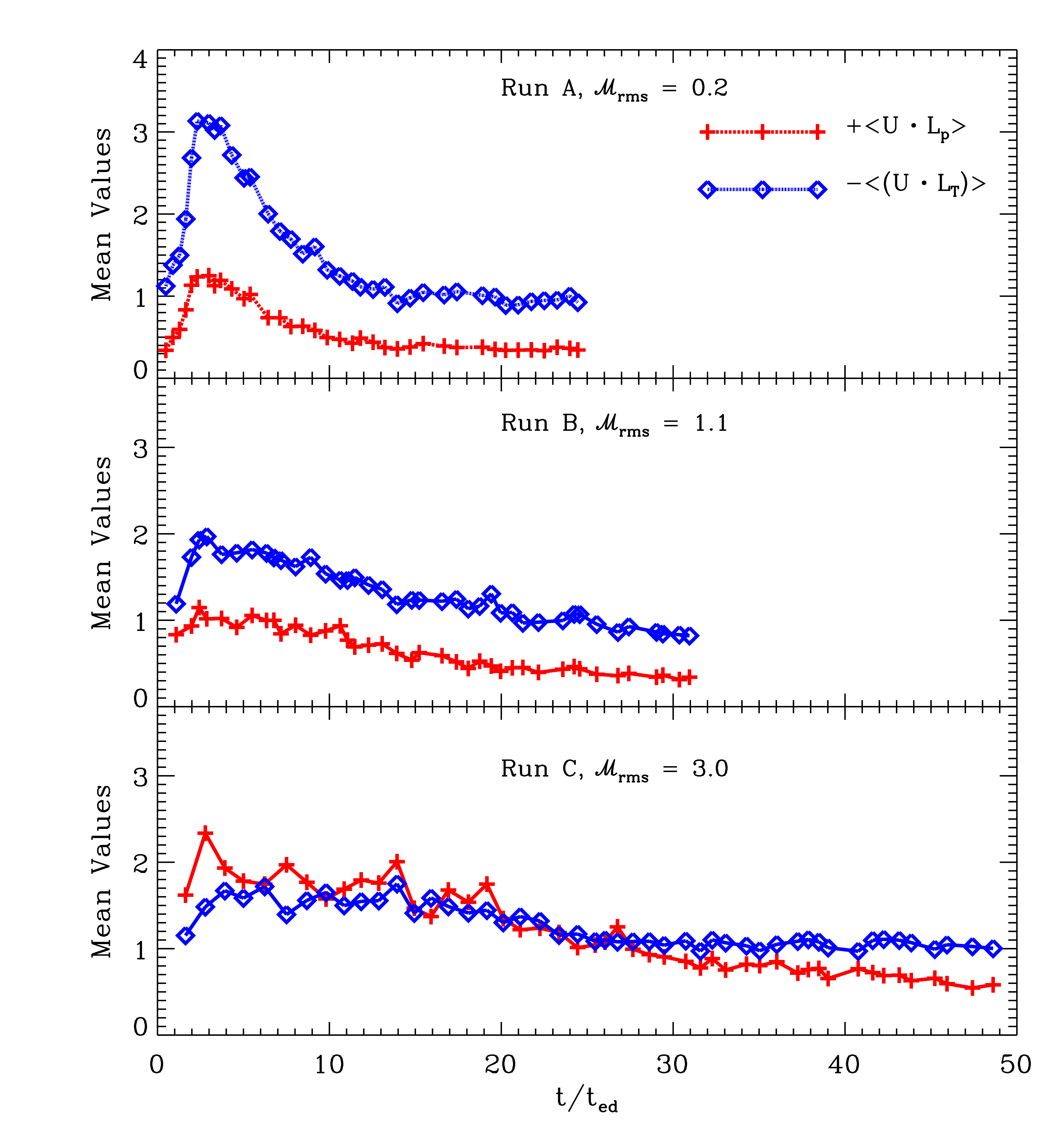} 
\caption{Evolution of the first and second term in equation~\ref{tevol_emag2} 
for runs A (top panel), B (middle panel), and C (bottom panel).
}
\label{fig:lforce_work}
\end{figure}

Using the instantaneous values of $\urms$ and $\brms$ for the normalisation 
parameter $\ted/\brms$, we show in Fig.~\ref{fig:lforce_work} the time evolution 
of the average values of the first term (blue dotted line with diamonds), and the 
second term (red, dashed line with '+' symbol). The fluctuations in the velocity 
are within $10$ per cent of the rms value. In all the three cases corresponding to 
different degrees of flow compressibility, both the terms have positive average 
values which implies that they contribute to the growth in magnetic energy (see 
equation~\ref{tevol_emag2}). In the subsonic and transonic cases (top and middle 
panel), the work done against the gradient of the magnetic pressure is always 
small compared to the magnetic tension. The plots also show that both magnetic 
pressure and magnetic tension are involved in dynamo saturation. We see that 
the average values of both the terms decrease as the dynamo evolved from the 
kinematic to the saturated phase. In the subsonic case, both magnetic pressure 
and tension contributions decrease by identical factors. However, in the transonic 
case, the suppression of the magnetic pressure term is more compared to the 
decrease due to the magnetic tension. 

Interestingly in the supersonic case (bottom panel), the energy gained 
from the velocity due to $\left\langle\UU\cdot \LL_{\rm T}\right\rangle$ and 
$\left\langle\UU\cdot \LL_{\rm p}\right\rangle$ terms are comparable up 
to $22\ted$. As can be seen from Fig.~\ref{fig:emag_evol}, this corresponds 
to the time by when the kinematic growth phase of the magnetic energy 
culminates and the intermediate phase of growth begins. Similar to the 
transonic case, once the dynamo evolves to the saturated phase, the work 
done against the gradient of the magnetic pressure is suppressed more in 
comparison to the magnetic tension. This again reflects the fact that the 
magnetic pressure gains in importance in opposing compressive motions 
as the dynamo saturates.

Thus, expressing the equation for magnetic energy evolution in terms of 
the rate of work done against the components of the Lorentz force offers 
a fresh perspective which highlights the significance of magnetic pressure 
in the supersonic case.

\section{Impact of local stretching, advection, compression, and dissipation 
on magnetic energy evolution} 
\label{local_effects}

Another approach to understand how the FD saturates is to analyse the 
effects of local stretching, advection, compression, and dissipation on the 
growth or decay of magnetic energy \citep{Seta+20, SF21b}. In this context, 
it is important to clarify if suppression of field line stretching is the only 
agent responsible for saturating the dynamo. To this end, we take the dot 
product of the magnetic induction equation with $\BB$ and then expand 
the $\BB\cdot[\nabla\times(\UU\times\BB)]$ term to obtain the evolution 
of the mean magnetic energy. In the following, we present a detailed 
derivation of the magnetic energy evolution equation and also highlight 
a subtle point. 

Assuming $\eta$ to be constant, the $i$-th component of the magnetic 
induction equation is
\begin{align}
\frac{\partial B_{i}}{\partial t} = [\nabla \times (\UU\times \BB)]_{i} - \eta(\nabla\times \JJ)_{i}, 
\end{align} 
where $\JJ = \nabla \times\BB$ is the current density with $\mu_{0} = 1$.

Expanding the cross-product in the induction term using Levi-Civita symbols 
yield
\begin{align}
\frac{\partial B_{i}}{\partial t} = \epsilon_{ijk}\frac{\partial}{\partial x_{j}}(\epsilon_{klm}U_{l}B_{m}) 
- \eta(\nabla\times\JJ)_{i}.
\end{align}

Now, substituting $\epsilon_{ijk}\epsilon_{klm} = (\delta_{il}\delta_{jm} - \delta_{im}\delta_{jl})$ 
and utilising the properties of Kronecker Delta $\delta_{ij}$'s gives, 
\begin{align}
\frac{\partial B_{i}}{\partial t} = \frac{\partial}{\partial x_{j}}(U_{i}B_{j}) - \frac{\partial}{\partial x_{j}}(U_{j}B_{i}) 
-\eta(\nabla\times\JJ)_{i}
\label{compb}
\end{align}

Taking a dot product of equation~(\ref{compb}) with $B_{i}$, and neglecting the 
term $\propto \nabla\cdot\BB$ we obtain the evolution equation of the magnetic 
energy in terms of local stretching, advection, compression, and dissipation 
terms, 

\begin{align}
\frac{\partial}{\partial t}\left(\frac{B^{2}}{2}\right) = \underbrace{B_{i}B_{j}\frac{\partial U_{i}}{\partial x_{j}}}_\text{stretching} 
- \underbrace{B_{i}\frac{\partial U_{j}}{\partial x_{j}}B_{i}}_\text{compression} 
- \underbrace{B_{i}U_{j}\frac{\partial B_{i}}{\partial x_{j}}}_\text{advection} 
- \underbrace{\eta B_{i}(\nabla\times\JJ)_{i}}_\text{dissipation}. 
\label{emevol_terms}
\end{align}

Next, we define the rate of the strain tensor $S_{ij} = (U_{i,j} + U_{j,i})/2$ 
where $U_{i,j}$ denote partial derivative of $U_{i}$ w.r.t. $j$ and vice versa. 
Now, expressing the stretching term in equation~(\ref{emevol_terms}) in terms of 
$S_{ij}$ and recasting the third term 
$B_{i}(U_{j}\partial_{j})B_{i}$ as $U_{j}\partial_{j}(B_{i}^{2}/2)$ the volume 
integrated magnetic energy evolution is given by

\begin{align}
\int_V \frac{1}{2} \frac{\partial |\BB|^2}{\partial t}~\dd V & = 
\int_{V}B_{i}B_{j}S_{ij}~\dd V - \int_{V}\UU\cdot\frac{1}{2}\nabla|\BB|^{2}~\dd V \nonumber \\
& - \int_{V}|\BB|^{2}(\nabla\cdot\UU)~\dd V \nonumber \\
& - \eta\int_{V}\BB\cdot(\nabla\times\JJ)~\dd V.
\label{emevol}
\end{align}

Note that in deriving equation~(\ref{emevol}), the term $\propto (U_{i,j} - U_{j,i})$ 
is anti-symmetric in $(i,j)$ and hence does not contribute to the magnetic 
energy. 

The first term on the right-hand side of equation~(\ref{emevol}) represents 
the stretching and compression of the magnetic field lines by the turbulent 
flow which may increase or decrease the magnetic energy. The second term 
reflects the effects of advection of the field lines by the turbulent flow. The third 
term representing the effects of compression can also reduce the magnetic 
energy depending on the divergence of the velocity field and is generally 
important in transonic and supersonic flows. The last term is dissipative in 
nature which acts to reduce the magnetic energy. The effects of local 
stretching, advection, compression, and dissipation on the magnetic 
energy growth (or decay) can then be analysed from the PDFs of each 
of the terms on the right-hand side of equation~(\ref{emevol}).

Before proceeding further, we wish to elucidate a subtle point. We note 
that the advection term in equation~(\ref{emevol}) can be further simplified to 
$[\nabla\cdot(\UU|\BB^{2}|) -|\BB^{2}|(\nabla\cdot\UU)]/2$ which together 
with the compression term leads to a term $\propto (\nabla\cdot\UU)$ 
and a surface term $\propto (\UU|\BB^{2}|)$. While for periodic boundary 
conditions, the latter term vanishes, it will still contribute when computing 
the PDFs at each mesh point. Similar reasoning is applicable to the 
dissipation term in equation~(\ref{emevol}) which can be expressed as a 
combination of a divergence term $\propto \nabla\cdot(\JJ\times\BB)$ and 
a term $\propto |\JJ|^{2}$. 
Again, the divergence term vanishes for periodic boundaries but will 
contribute to the PDFs at each mesh point. Therefore, while computing 
the PDFs it is essential to retain the contribution from the advection and 
dissipation terms in the form given in equation~(\ref{emevol}). 
This is in contrast to \cite{SF21b}, where the aforesaid surface terms 
were neglected. 

Similar to in Section~\ref{impact} we express equation~(\ref{emevol}) in dimensionless 
form by multiplying both sides of the equation by $t_{\rm ed}/B_{\rm rms}^{2}$ 
which results in
\begin{align}
\int_{V}\frac{\partial}{\partial t}\left(\frac{|\BB|^{2}}{2}\right)\,\frac{t_{\rm ed}}{B_{\rm rms}^{2}}~\dd V &=
+ \int_{V}S_{ij}B_{i}B_{j}\frac{\ted}{B_{\rm rms}^{2}}~\dd V \nonumber \\
&- \int_{V}\UU\cdot\frac{1}{2}\nabla|\BB|^{2}\frac{\ted}{\brms^{2}}~\dd V \nonumber \\
&- \int_{V}|{\BB}|^{2}({\nabla\cdot\UU})\frac{\ted}{B_{\rm rms}^{2}}~\dd V \nonumber \\
&- \eta\int_{V}\BB\cdot(\nabla\times\JJ)\frac{t_{\rm ed}}{B_{\rm rms}^{2}}~\dd V.
\label{emevol_dimless}
\end{align}

While computing the PDFs of the compression, advection, and dissipation 
terms are fairly straightforward, to compute the PDFs of the local stretching 
we first project the local magnetic field along the three eigenvectors of the 
rate of strain tensor ($\ee_{1}, \ee_{2}, \ee_{3}$). The corresponding 
eigenvalues are denoted as $\lambda_{1} > \lambda_{2} > \lambda_{3}$, 
respectively \citep{B95,SPS14b,Seta+20,SF21b}. 
In the incompressible case, $\lambda_{1} + \lambda_{2} + \lambda_{3} = 0$ 
while in the supersonic case the sum of the three eigenvalues need not 
be zero. In general, $\ee_{1}$ and $\ee_{3}$ correspond to the directions 
of local stretching and compression while $\ee_{2}$ can be either, depending 
on sign of $\lambda_{2}$ \citep{Ash+87}.

We then compute the local stretching as, 
$\Sigma_{i=1,2,3}[\lambda_{i}\ted(\BB\cdot\ee_{i})^{2}]/\brms^{2}$, 
where ($\BB\cdot\ee_{i}$) denotes the component of the local magnetic 
field along the direction of the eigenvectors, and $\lambda_{i}$ are the 
eigenvalues corresponding to the eigenvectors $\ee_{i}$. 

\begin{figure}
\centering
\includegraphics[width=\columnwidth]{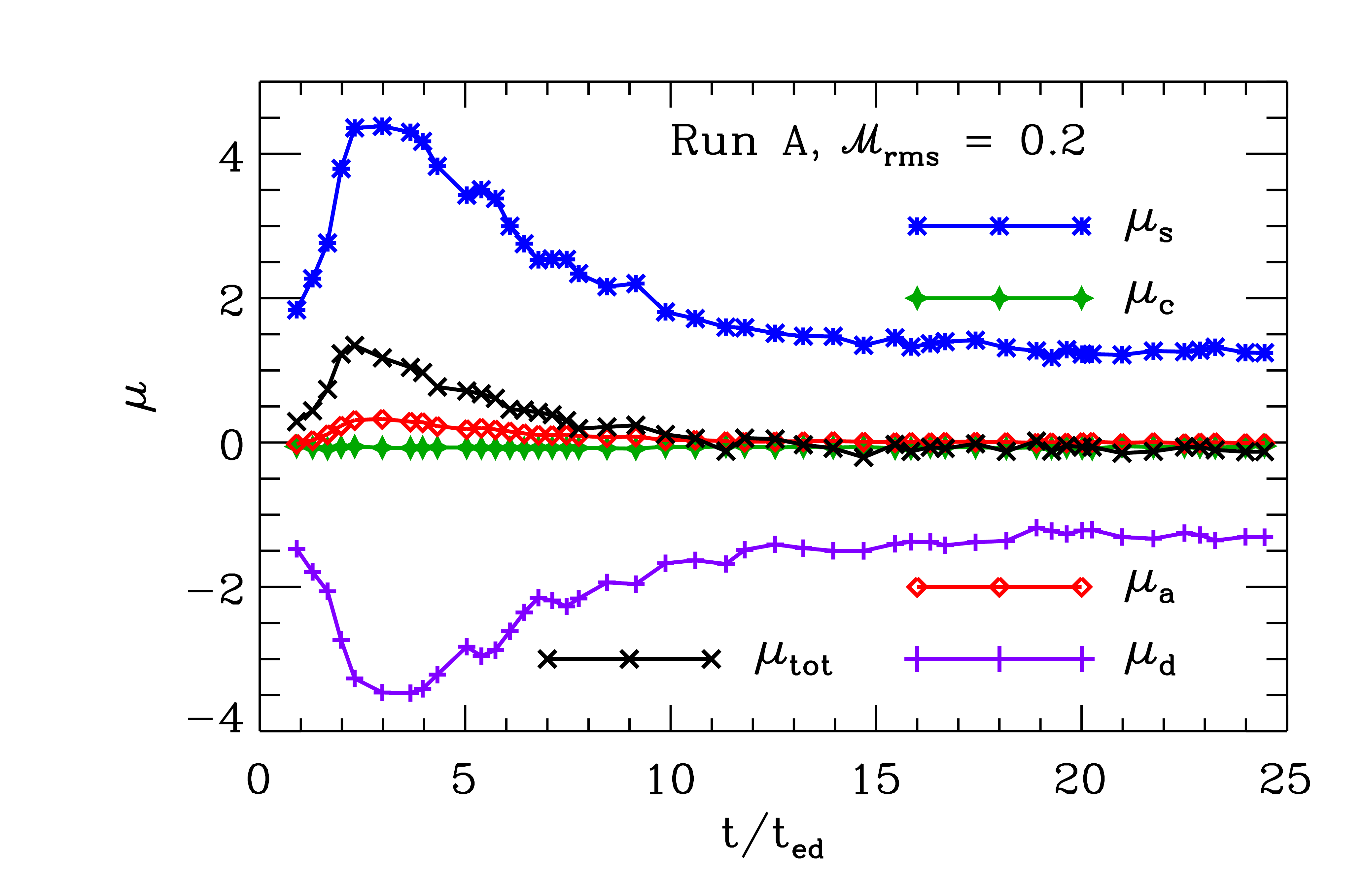} 
\caption{Evolution of the mean values ($\mu$) of the PDFs of local 
stretching (blue, asterisks), compression (green, filled stars), advection (red, 
open diamonds) and dissipation term (purple, '+') for the subsonic 
run with $\mrms = 0.2$. The sum total of the mean values is depicted by 
black crosses. 
}
\label{fig:pdf_mean_sub}
\end{figure}

\begin{figure}
\centering
\includegraphics[width=\columnwidth]{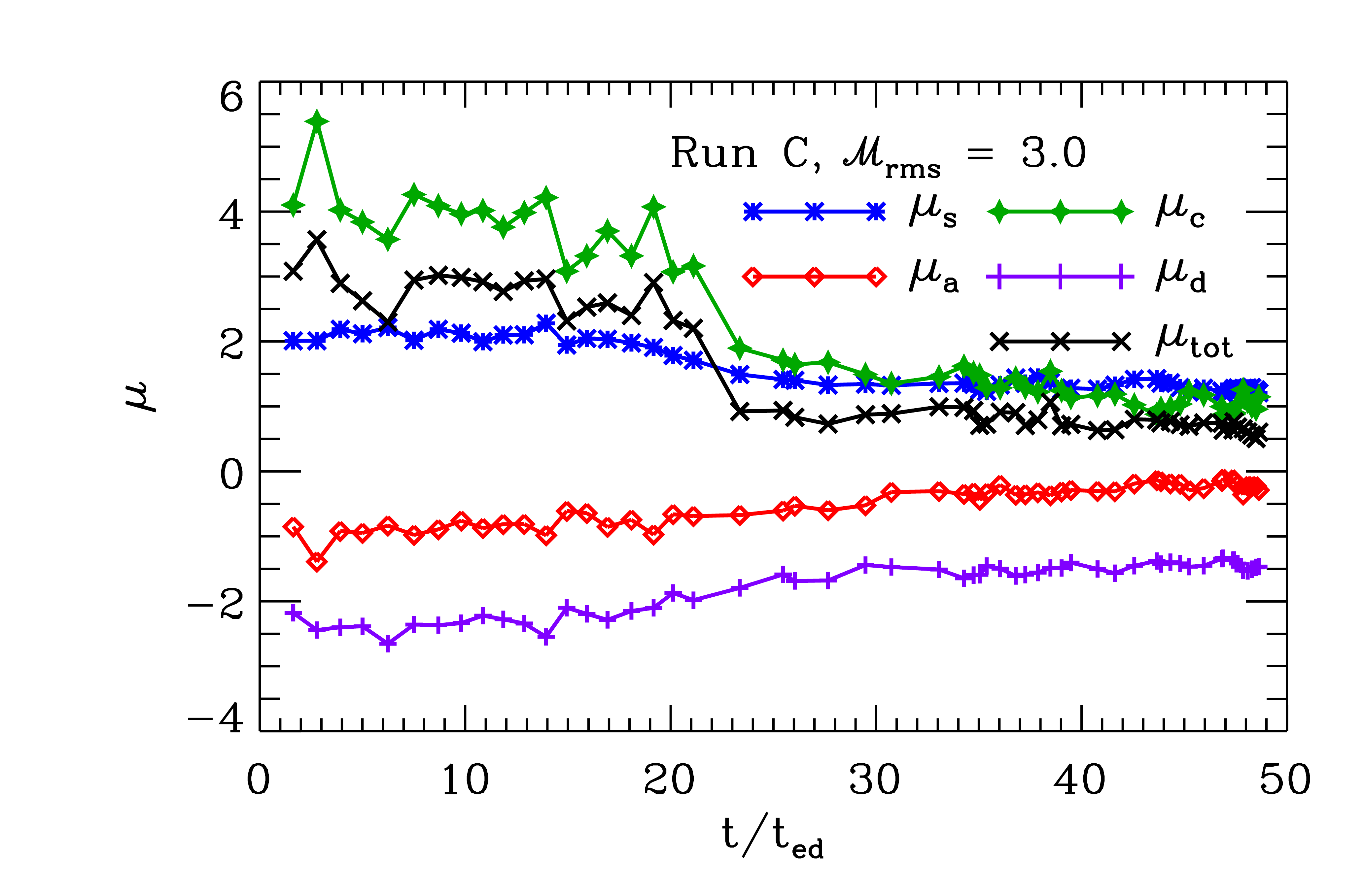} 
\caption{Same as in Fig.~\ref{fig:pdf_mean_sub} but now for the 
supersonic run with $\mrms = 3.0$.}
\label{fig:pdf_mean_super}
\end{figure}

In the kinematic phase, there is a net growth in the magnetic energy, while 
in the saturated phase the magnetic energy is expected to remain constant. 
It is expected that this is caused by the mutual adjustments between the 
local stretching, compression, advection, and dissipation terms. In what 
follows, we discuss the effects of each of these terms in two regimes of 
compressibility focussing on run A where $\mrms = 0.2$ and run C where 
$\mrms = 3.0$. Since we normalised equation~(\ref{emevol_dimless}) using the 
instantaneous values of $\urms$ and $\brms$, we refrain from computing 
the total PDFs over the range of $\ted$ in the kinematic and saturated 
phases. Instead, we show in Fig.~\ref{fig:pdf_mean_sub} the time evolution 
of the mean values of the PDFs of local stretching ($\mu_{\rm s}$), advection 
($\mu_{\rm a}$), compression ($\mu_{\rm c}$), and dissipation ($\mu_{\rm d}$) 
and the sum total of the mean values ($\mu_{\rm tot}$) for the subsonic 
run with $\mrms = 0.2$. In the same vein, Fig.~\ref{fig:pdf_mean_super} 
shows the evolution for the supersonic run with $\mrms = 3$. For the sake 
of completeness, the PDFs of stretching, advection, compression, and 
dissipation terms at a few representative times in the kinematic and 
saturated phases in shown in Figs~\ref{fig:pdfs_emag_sub} and 
\ref{fig:pdfs_emag_super}.

In the subsonic case (run A), Fig.~\ref{fig:pdf_mean_sub} shows that field 
amplification in the kinematic phase is predominantly driven by an increase 
in local stretching of the field by the turbulent flow. This is accompanied by 
an increase in dissipation. The mean values of local stretching, advection,
and dissipation peaks $\approx 3-4\ted$ beyond which, the growth rate of the 
magnetic energy changes (see Fig.~\ref{fig:emag_evol}). In fact, 
Fig.~\ref{fig:pdfs_emag_sub} show that dissipation is relatively stronger than 
both field line stretching and advection initially. 
Beyond this, the change in growth rate of magnetic energy is manifested 
by a gradual decrease in $\mu_{\rm s}, \mu_{\rm a}$ and $\mu_{\rm d}$. This 
continues up to $t/\ted \approx 15$ beyond which, all three attain steady state 
values. However, $\mu_{\rm c}$ remains negligible throughout the evolution. 
This reconfirms the fact that compression plays a negligible role in dynamo 
saturation in subsonic flows. The net growth of magnetic energy in the kinematic 
phase is confirmed from the non-zero value of the total mean. Of course, as 
the dynamo approaches the non-linear phase, the net growth consistently 
decreases. In the saturated phase, field line stretching and dissipation 
compensate each other (i.e., $|\mu_{\rm s}| \approx |\mu_{\rm d}|$) while 
$\mu_{\rm a}$ is totally suppressed to negligible values. 
A quick comparison between the PDFs in Fig.~\ref{fig:pdfs_emag_sub} show 
that both local stretching and dissipation are reduced in the saturated phase. 
Note that the rate of strain tensor in the stretching term in equation~(\ref{emevol_dimless})
is only affected by magnetic tension. Thus, the reduction in stretching observed 
in the saturated phase suggest that magnetic tension is resisting the further 
stretching of the field lines. 

In contrast to the subsonic case, Fig.~\ref{fig:pdf_mean_super} shows 
that amplification of the magnetic field in the kinematic phase in supersonic 
turbulence (run C) is dominated by compression of magnetic field lines. 
Growth of magnetic energy due to stretching of the field lines remains 
subdominant to the effects of compression by almost a factor of two in this 
phase. The compressible nature of turbulence ($\mrms = 3.0$) drives local 
density enhancements which further amplifies the magnetic field. On the 
other hand, the effects of advection remain marginal compared to 
compression, stretching, and dissipation. 
As the magnetic energy transitions to the intermediate phase of growth, 
both compression and field line stretching reduces marked by a drop in 
$\mu_{\rm s}$ and $\mu_{\rm c}$ after $t/\ted \approx 22$. At the same 
time, advection and dissipation also decreases slightly.
In the saturated phase, $\mu_{\rm s}$ and $\mu_{\rm c}$ become 
comparable to each other. The overall evolution of the mean values 
spanning from kinematic to saturated phase show that field amplification 
by compressive motions suffer the strongest suppression as $\mu_{\rm c}$ 
decreases by a factor $\approx 3.3$ (also see Fig.~\ref{fig:pdfs_emag_super}) 
compared to a factor of $\approx 1.5$ for the reduction in stretching. This 
implies that in the compressible regime, in addition to reduction in field line 
stretching, advection, and dissipation, the suppression of further compression 
of field lines due to the action of magnetic pressure plays a major role in 
saturating the dynamo. Interestingly, the aforesaid deduction is consistent 
with \citet{SPS14b} where it was shown that compressive motions of the 
strain (perpendicular to the direction of stretching) are suppressed in the 
saturated phase of the FD. 

\section{Conclusions}
\label{conclu}

Over the course of the last two decades, a great deal of progress has been 
made to understand the manifold features of FDs and it's 
application to magnetic fields in the ISM of galaxies and in the intracluster 
medium of galaxy clusters. A majority of these studies were based 
on direct numerical simulations that dealt with amplification of dynamically 
insignificant seed magnetic fields and subsequent saturation in flows
encompassing various degrees of compressibility. However, the precise 
way in which the dynamo evolves to the non-linear saturated state and the 
factors that play a deciding role in achieving this steady state remained 
far from understood, more so in compressible flows. While in a broader 
sense, the back reaction due to the Lorentz force is rightly adjudged to be 
responsible for saturating the dynamo, it had been argued that the 
magnetic tension force is the main ingredient that suppresses Lagrangian 
chaos via reduction of field line stretching which in turns results in the 
saturation of the dynamo \citep{CHK96, R19} (also see \citet{E11} 
and \citet{ELV11}).

Although the role of magnetic tension is appreciated for dynamo saturation 
in incompressible flows, a better understanding is required in compressible 
flows which are likely to occur in the galactic ISM where density fluctuations 
can be significant. In particular, two important questions arise - does magnetic 
pressure gradient associated with the Lorentz force also play a role in 
saturating the dynamo and, is suppression of field line stretching the only 
mechanism responsible for dynamo saturation when the condition of strict 
incompressibility is relaxed? 

With the aforesaid questions in mind, here we have provided a new perspective 
on how FDs saturate. To this effect, we performed numerical simulations of FDs 
at $\Pm = 1$ in turbulent flows driven at half the scale of the box with the amplitude 
of turbulent driving adjusted such that the steady state $\mrms = 0.2, 1.1$ and 
$3.0$. We then analysed the data from these simulations using a variety of probes. 
These include : (i) 
exploring the correlation between the density and magnetic field strength, 
(ii) the nature of the alignments between the magnetic field, the gradient 
of the density, and the components of the Lorentz force, (iii) the work done 
against different parts of the Lorentz force to grow the magnetic field, and 
(iv) the impact of local stretching, advection, compression, and dissipation 
on the evolution of the magnetic energy. Our main conclusions are as follows : 

\begin{enumerate} 

\item The PDFs of the magnetic energy density shown in Fig.~\ref{fig:pdfs_bmag}, 
demonstrate in a novel manner that together with the general sea of less 
intense, volume filling fields, strong fields regions are also likely to play a role 
in dynamo saturation in supersonic turbulence. In fact, as Fig.~\ref{fig:pdfs_bmag} 
shows strong fields of strength $\approx 10\brms$ in the supersonic case 
contribute comparable energy density as the rms strength fields in the 
kinematic phase. Even after saturation, fields upto $4\brms$ contribute 
significantly to the magnetic energy density.

\item Density and magnetic fields strength are anticorrelated in 
subsonic flows. Due to the incompressible nature of the flow, densities 
cannot be changed by a large factor. Thus, saturation of the dynamo 
can happen by suppression of field line stretching. This is also confirmed 
from Fig.~\ref{fig:pdf_mean_sub}. 

\item As the effects of compressibility become important (e.g. runs B 
and C) magnetic fields are amplified by a combination of random 
stretching and compression. 
The latter ensures an initial positive value of $r_{p}$ which then 
decreases as dynamo saturates (see Fig.~\ref{fig:corr_sub_trans_super}). 
In transonic flows, $\rho$ and $B$ tend to become uncorrelated in the 
saturated state while in the supersonic case, $r_{p}$ shows a lesser 
degree of positive correlation in the saturated phase compared to the 
value in the kinematic phase. 

\item A quick glance at Fig.~\ref{fig:bcuts} reveals that the anti-correlation 
between $\rho$ and $B$ seen in the subsonic case stems from the strong 
field regions with $B/\brms > 3$. Such rarer and intense fields also lead to 
negligible $r_{p}$ in the transonic case and a much weaker positive 
correlation in the supersonic case. Interestingly, we find that regions with 
$\delta\meanrho/\meanrho < \mathcal{M}^{2}$ (see Fig.~\ref{fig:dcuts}) 
lead to the anticorrelation in the subsonic case. In the transonic case, 
these over dense regions tend to be uncorrelated with the magnetic field 
while in the supersonic case, they tend to reduce the degree of positive 
correlation. 

\item The fact that magnetic pressure forces play a significant role in 
dynamo saturation is clearly evident from the PDFs of the alignment 
presented in Fig.~\ref{fig:pdf_sub_super}. In the subsonic case, we find 
that ${\bf n}_{\nabla\rho}$ is statistically aligned anti-parallel to 
${\bf n}_{\nabla B^{2}}$, with the alignment becoming stronger in the 
saturated phase. In the transonic and more prominently in the 
supersonic case, apart from the antiparallel alignment, a strong 
parallel alignment between the two emerges in the kinematic phase. 
Nevertheless, the antiparallel alignment due to magnetic pressure 
forces dominate in the saturated phase. In contrast, magnetic tension 
does not show any preferred alignment with the gradient of the density 
irrespective of the flow compressibility. 

\item An analysis of the impact of Lorentz forces on the growth of 
magnetic energy (Fig.~\ref{fig:lforce_work}) reveals that the contribution 
from the magnetic pressure gradient is always small compared to the 
magnetic tension up to $\mrms = 1.1$. However, in the supersonic 
regime, the two are comparable till the time the non-linear saturated 
phase ushers in. Thereafter, the work done against the gradient of 
magnetic pressure (aiding field growth) is suppressed more in 
comparison to the magnetic tension. This again illustrates the 
importance of magnetic pressure gradients in dynamo saturation for 
supersonic turbulence.

\item The time evolution of the mean values of the PDFs of 
local stretching, advection, compression, and dissipation terms 
are shown in Figs~\ref{fig:pdf_mean_sub} and \ref{fig:pdf_mean_super}. 
They clearly demonstrates firstly that local stretching of the 
field lines play a crucial role in amplifying the field in subsonic flows.
And the saturation of the dynamo in such flows occurs via a 
reduction in field line stretching. In contrast, in supersonic flows, 
amplification of magnetic fields is dominated by compression of 
the field lines in the kinematic phase with stretching playing a 
subdominant role. As the magnetic energy evolves to equipartition 
strengths, the dynamo saturates more due to suppression of 
compressive amplification than due reduction in field line stretching.

\end{enumerate}

In a nutshell, the results from our work confirm in detail a new route 
to dynamo saturation based on the simple schematic picture of magnetic 
flux tubes discussed in Section~\ref{intro}. In incompressible flows,
where magnetic tension plays a decisive role in dynamo saturation, 
magnetic pressure forces nevertheless act simultaneously to oppose 
the compression of the flux tubes, especially in rare but intense 
magnetic structures. The pivotal role of magnetic pressure forces in 
saturating the dynamo comes into prominence in transonic and 
particularly in supersonic flows.There, growing magnetic pressure 
forces resist compression of flux tubes and lead to regions of strong 
magnetic flux being emptied out rather than a decrease of their 
cross-sectional areas. This fact has not been sufficiently appreciated 
so far and could be important as many astrophysical systems host
compressible turbulent flows. Future work extending these considerations 
also to cases with $\Pm \gg 1$, and where turbulence is inhomogeneous 
as expected in galaxies, would be of interest.

FDs play an important role in amplifying weak seed magnetic 
fields to strengths of dynamical importance. In astrophysical systems like 
galaxies, where conditions are favourable for large-scale dynamo action, the 
fields amplified by FDs provide a seed for the later evolution 
\citep{SS21}. Here, turbulence is mainly driven by supernovae explosions. 
On the other hand, during the formation of the First stars FD
action can be triggered due to gravitational collapse which generates turbulent 
motions \citep{Sur+10, Sur+12,FSSBK11}. On even larger scales, the same 
mechanism may amplify fields during the formation of the first galaxies 
\citep{Latif+13,SSK13,Pak+17} and also in galaxy clusters \citep{SSH06,CR09, 
Vazza+18}. In the latter, such dynamo generated fields crucially impact the 
properties of polarized synchrotron emission \citep{BS21,SBS21}. Thus, 
elucidating the role of the magnetic pressure and magnetic tension in dynamo 
saturation is of lasting importance.

\section*{Acknowledgements}

SS thanks the Science and Engineering Research Board (SERB) of 
the Department of Science \& Technology (DST), Government of India, 
for support through research grant ECR/2017/001535. He also acknowledges 
computing time awarded at the CDAC National Param Supercomputing 
facility (NPSF), India, under the grant ‘Hydromagnetic-Turbulence-PR’ 
and the use of the High Performance Computing (HPC) resources made 
available by the Computer Centre of the Indian Institute of Astrophysics (IIA).
The authors thank the referee for a detailed and constructive report. 
The software used in this work was developed in part by the DOE 
NNSA-and DOE Office of Science supported Flash Center for Computational 
Science at the University of Chicago and the University of Rochester.

\section*{Data Availability}

The data supporting this article are available on reasonable request to the corresponding 
author.



\appendix

\section{Evolution of the kinetic and magnetic spectra}
\label{evol_spec} 

\begin{figure*}
\centering
\includegraphics[width=\textwidth]{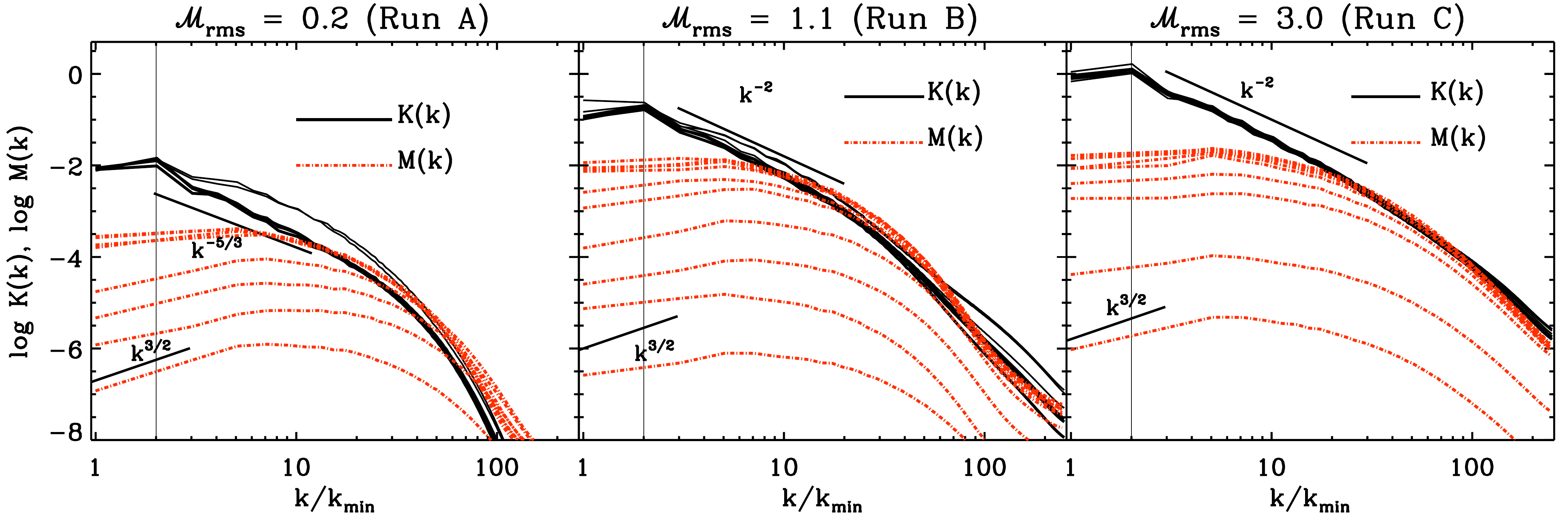} 
\caption{Time evolution of the kinetic $K(k)$, and magnetic $M(k)$ 
energy spectra in runs A (left panel), B (middle panel), and C (right panel). 
$K(k)$ is shown by solid, black lines while $M(k)$ is denoted by red, 
dash-dotted lines. The wave number is normalised to $k_{\rm min} = 2\pi$. 
The thin vertical line in each panel shows the turbulent forcing 
wave number.}
\label{fig:spec_comp}
\end{figure*}

In Fig.~\ref{fig:spec_comp}, we show the time evolution of the kinetic $K(k)$ and 
magnetic $M(k)$ energy spectra as a function of $k/k_{\rm min}$ for the three 
simulations presented here. $K(k)$ in the kinematic and intermediate phases 
are shown by thin, black solid line and in the saturated phase by thick, black solid 
lines. On the other hand, red dashed dotted curves depict the evolution of $M(k)$.

The figure shows that $K(k)$ develop a significant inertial range in all three runs 
whose slope is in reasonable agreement with a $k^{-5/3}$ for run A (subsonic) 
and $k^{-2}$ in runs B (transonic) and C (supersonic). Thin solid lines with power-law 
behaviour of the form $k^{3/2}$ is also shown in each panel for comparison 
with the Kazantsev spectra for magnetic energy \citep{K68}. The $M(k)$ 
shows the familiar self-similar evolution in the growth phase.  Later, as the 
dynamo saturates the spectra of $M(k)$ bunch together. We find that this 
happens from $t/\ted = 15, 22$ and $t/\ted = 30$ in runs A, B and C, respectively.

\section{Evolution of the PDFs of stretching, advection, compression and dissipation in 
compressible flows} 
\label{pdfs}

\begin{figure*}
\centering
\includegraphics[width=\textwidth]{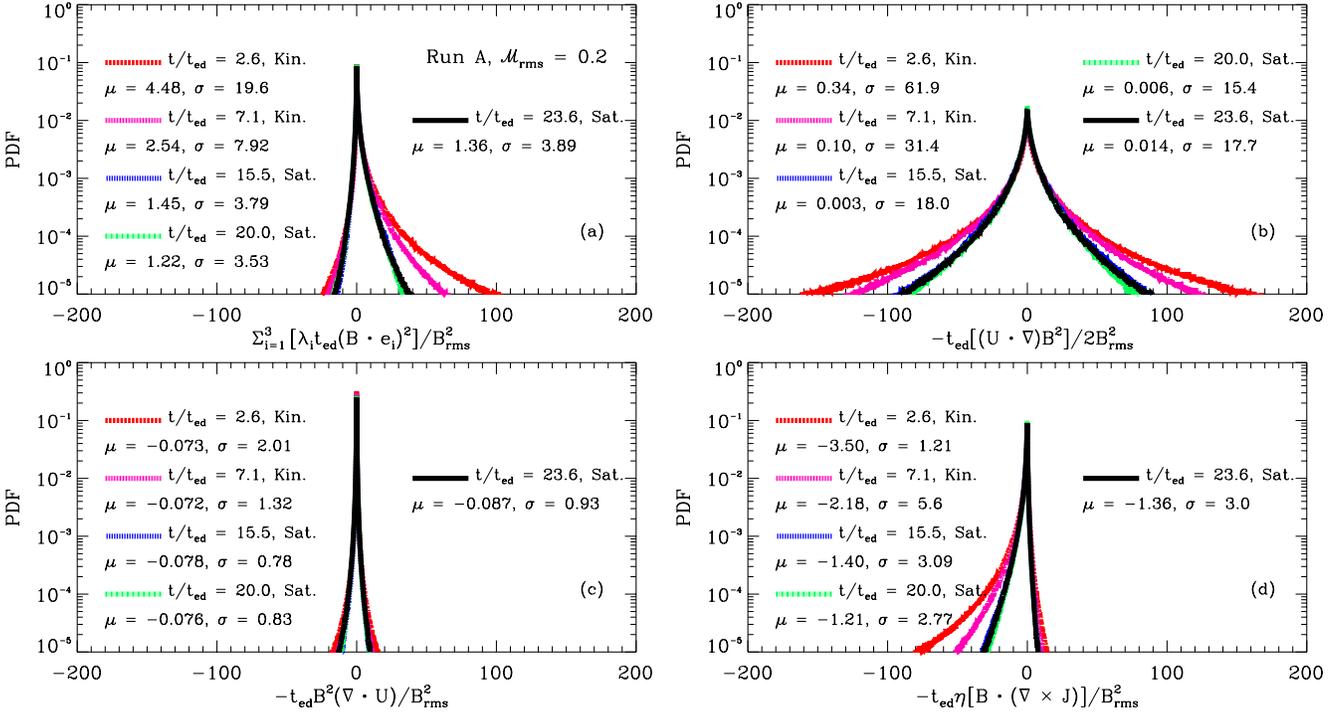} 
\caption{Evolution of the PDFs of the stretching term in panel (a), the 
advection term in panel (b), the compression term in panel (c) and 
the dissipation term in panel (d) for run A with $\mrms = 0.2$ (subsonic). 
The values of $\mu$ and $\sigma$ shown in the legends corresponds to 
the mean and standard deviation of the distribution in each case.}
\label{fig:pdfs_emag_sub}
\end{figure*}

\begin{figure*}
\centering
\includegraphics[width=\textwidth]{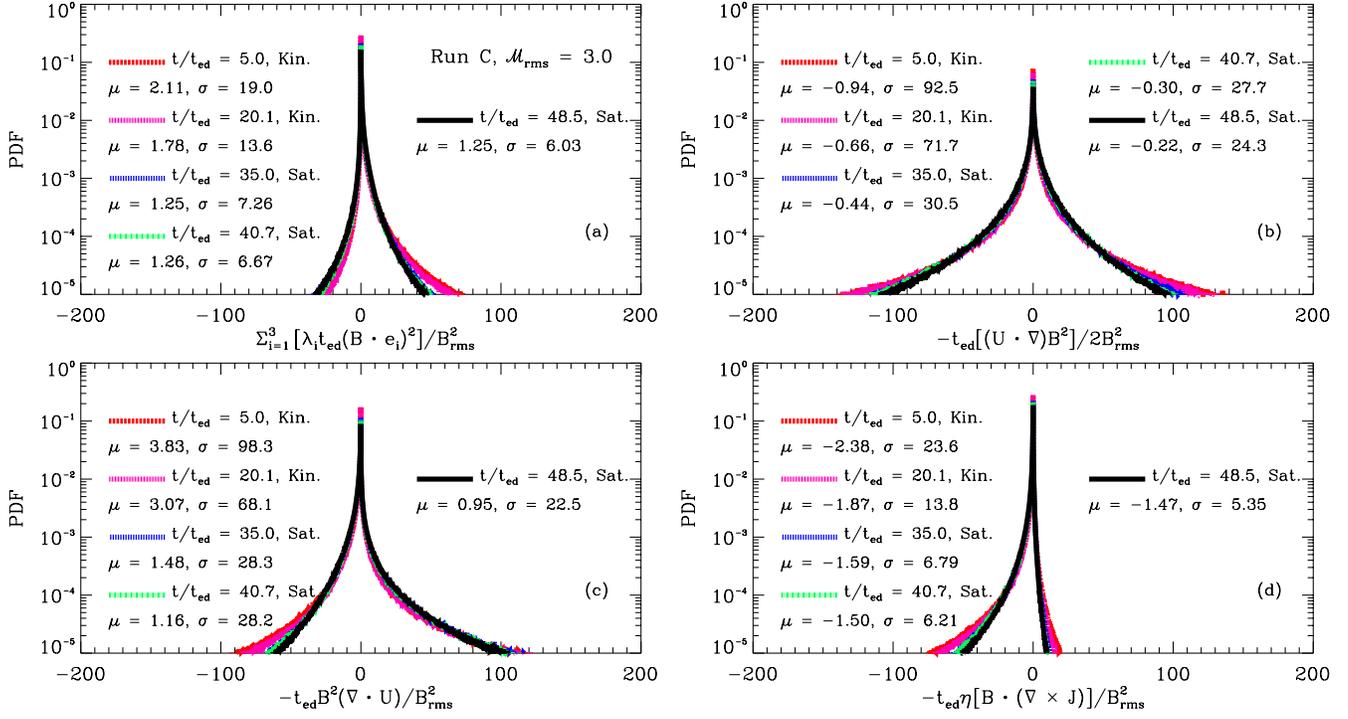} 
\caption{Same as in Fig.~\ref{fig:pdfs_emag_sub} but now 
for run C with $\mrms = 3.0$ (supersonic).}
\label{fig:pdfs_emag_super}
\end{figure*}

In Figs~\ref{fig:pdfs_emag_sub} and \ref{fig:pdfs_emag_super} we show the 
PDFs of local stretching (in panel 'a'), advection (in panel 'b'), compression (in panel 'c'),
and dissipation (in panel 'd') at five representative times encompassing the 
kinematic and saturated phases of the dynamo. It is apparent from these plots 
that the standard deviations ($\sigma$) values are much larger compared to 
the mean ($\mu$) values. Fig.~\ref{fig:pdfs_emag_sub} clearly shows that the 
compression term plays a negligible role in turbulent incompressible flows. 
On the other hand, we can see from Fig.~\ref{fig:pdfs_emag_super} that 
it does play an important role is supersonic flows.

\bsp	
\label{lastpage}

\begin{thebibliography}{}
\makeatletter
\relax
\def\mn@urlcharsother{\let\do\@makeother \do\$\do\&\do\#\do\^\do\_\do\%\do\~}
\def\mn@doi{\begingroup\mn@urlcharsother \@ifnextchar [ {\mn@doi@}
  {\mn@doi@[]}}
\def\mn@doi@[#1]#2{\def\@tempa{#1}\ifx\@tempa\@empty \href
  {http://dx.doi.org/#2} {doi:#2}\else \href {http://dx.doi.org/#2} {#1}\fi
  \endgroup}
\def\mn@eprint#1#2{\mn@eprint@#1:#2::\@nil}
\def\mn@eprint@arXiv#1{\href {http://arxiv.org/abs/#1} {{\tt arXiv:#1}}}
\def\mn@eprint@dblp#1{\href {http://dblp.uni-trier.de/rec/bibtex/#1.xml}
  {dblp:#1}}
\def\mn@eprint@#1:#2:#3:#4\@nil{\def\@tempa {#1}\def\@tempb {#2}\def\@tempc
  {#3}\ifx \@tempc \@empty \let \@tempc \@tempb \let \@tempb \@tempa \fi \ifx
  \@tempb \@empty \def\@tempb {arXiv}\fi \@ifundefined
  {mn@eprint@\@tempb}{\@tempb:\@tempc}{\expandafter \expandafter \csname
  mn@eprint@\@tempb\endcsname \expandafter{\@tempc}}}

\bibitem[\protect\citeauthoryear{{Achikanath Chirakkara}, {Federrath},
  {Trivedi}  \& {Banerjee}}{{Achikanath Chirakkara} et~al.}{2021}]{CFTB21}
{Achikanath Chirakkara} R.,  {Federrath} C.,  {Trivedi} P.,   {Banerjee} R.,
  2021, \mn@doi [\prl] {10.1103/PhysRevLett.126.091103}, \href
  {https://ui.adsabs.harvard.edu/abs/2021PhRvL.126i1103A} {126, 091103}

\bibitem[\protect\citeauthoryear{{Ashurst}, {Kerstein}, {Kerr}  \&
  {Gibson}}{{Ashurst} et~al.}{1987}]{Ash+87}
{Ashurst} W.~T.,  {Kerstein} A.~R.,  {Kerr} R.~M.,   {Gibson} C.~H.,  1987,
  \mn@doi [Physics of Fluids] {10.1063/1.866513}, \href
  {https://ui.adsabs.harvard.edu/abs/1987PhFl...30.2343A} {30, 2343}

\bibitem[\protect\citeauthoryear{{Basu} \& {Sur}}{{Basu} \& {Sur}}{2021}]{BS21}
{Basu} A.,  {Sur} S.,  2021, \mn@doi [Galaxies] {10.3390/galaxies9030062},
  \href {https://ui.adsabs.harvard.edu/abs/2021Galax...9...62B} {9, 62}

\bibitem[\protect\citeauthoryear{{Batchelor}}{{Batchelor}}{1950}]{B50}
{Batchelor} G.~K.,  1950, \mn@doi [Proceedings of the Royal Society of London
  Series A] {10.1098/rspa.1950.0069}, \href
  {https://ui.adsabs.harvard.edu/abs/1950RSPSA.201..405B} {201, 405}

\bibitem[\protect\citeauthoryear{{Benzi}, {Biferale}, {Fisher}, {Kadanoff},
  {Lamb}  \& {Toschi}}{{Benzi} et~al.}{2008}]{Benzi+08}
{Benzi} R.,  {Biferale} L.,  {Fisher} R.~T.,  {Kadanoff} L.~P.,  {Lamb} D.~Q.,
   {Toschi} F.,  2008, \mn@doi [Physical Review Letters]
  {10.1103/PhysRevLett.100.234503}, \href
  {http://adsabs.harvard.edu/abs/2008PhRvL.100w4503B} {100, 234503}

\bibitem[\protect\citeauthoryear{{Bhat} \& {Subramanian}}{{Bhat} \&
  {Subramanian}}{2013}]{BS13}
{Bhat} P.,  {Subramanian} K.,  2013, \mn@doi [\mnras] {10.1093/mnras/sts516},
  \href {http://adsabs.harvard.edu/abs/2013MNRAS.429.2469B} {429, 2469}

\bibitem[\protect\citeauthoryear{{Brandenburg}}{{Brandenburg}}{1995}]{B95}
{Brandenburg} A.,  1995, \mn@doi [Chaos Solitons and Fractals]
  {10.1016/0960-0779(94)00177-R}, \href
  {https://ui.adsabs.harvard.edu/abs/1995CSF.....5.2023B} {5, 2023}

\bibitem[\protect\citeauthoryear{{Brandenburg} \& {Subramanian}}{{Brandenburg}
  \& {Subramanian}}{2005}]{BS05}
{Brandenburg} A.,  {Subramanian} K.,  2005, \mn@doi [\physrep]
  {10.1016/j.physrep.2005.06.005}, \href
  {http://adsabs.harvard.edu/abs/2005PhR...417....1B} {417, 1}

\bibitem[\protect\citeauthoryear{{Cattaneo}, {Hughes}  \& {Kim}}{{Cattaneo}
  et~al.}{1996}]{CHK96}
{Cattaneo} F.,  {Hughes} D.~W.,   {Kim} E.-J.,  1996, \mn@doi [\prl]
  {10.1103/PhysRevLett.76.2057}, \href
  {https://ui.adsabs.harvard.edu/abs/1996PhRvL..76.2057C} {76, 2057}

\bibitem[\protect\citeauthoryear{{Cho} \& {Ryu}}{{Cho} \& {Ryu}}{2009}]{CR09}
{Cho} J.,  {Ryu} D.,  2009, \mn@doi [\apjl] {10.1088/0004-637X/705/1/L90},
  \href {https://ui.adsabs.harvard.edu/abs/2009ApJ...705L..90C} {705, L90}

\bibitem[\protect\citeauthoryear{{Donnert}, {Vazza}, {Br{\"u}ggen}  \&
  {ZuHone}}{{Donnert} et~al.}{2018}]{Donnert+18}
{Donnert} J.,  {Vazza} F.,  {Br{\"u}ggen} M.,   {ZuHone} J.,  2018, \mn@doi
  [\ssr] {10.1007/s11214-018-0556-8}, \href
  {https://ui.adsabs.harvard.edu/abs/2018SSRv..214..122D} {214, 122}

\bibitem[\protect\citeauthoryear{{Eswaran} \& {Pope}}{{Eswaran} \&
  {Pope}}{1988}]{EP88}
{Eswaran} V.,  {Pope} S.~B.,  1988, \mn@doi [Physics of Fluids]
  {10.1063/1.866832}, \href {http://adsabs.harvard.edu/abs/1988PhFl...31..506E}
  {31, 506}

\bibitem[\protect\citeauthoryear{{Eyink}}{{Eyink}}{2011}]{E11}
{Eyink} G.~L.,  2011, \mn@doi [\pre] {10.1103/PhysRevE.83.056405}, \href
  {https://ui.adsabs.harvard.edu/abs/2011PhRvE..83e6405E} {83, 056405}

\bibitem[\protect\citeauthoryear{{Eyink}, {Lazarian}  \& {Vishniac}}{{Eyink}
  et~al.}{2011}]{ELV11}
{Eyink} G.~L.,  {Lazarian} A.,   {Vishniac} E.~T.,  2011, \mn@doi [\apj]
  {10.1088/0004-637X/743/1/51}, \href
  {https://ui.adsabs.harvard.edu/abs/2011ApJ...743...51E} {743, 51}

\bibitem[\protect\citeauthoryear{{Federrath}}{{Federrath}}{2016}]{Fed16}
{Federrath} C.,  2016, \mn@doi [Journal of Plasma Physics]
  {10.1017/S0022377816001069}, \href
  {http://adsabs.harvard.edu/abs/2016JPlPh..82f5301F} {82, 535820601}

\bibitem[\protect\citeauthoryear{{Federrath}, {Chabrier}, {Schober},
  {Banerjee}, {Klessen}  \& {Schleicher}}{{Federrath} et~al.}{2011a}]{Fed+11}
{Federrath} C.,  {Chabrier} G.,  {Schober} J.,  {Banerjee} R.,  {Klessen}
  R.~S.,   {Schleicher} D.~R.~G.,  2011a, \mn@doi [\prl]
  {10.1103/PhysRevLett.107.114504}, \href
  {http://adsabs.harvard.edu/abs/2011PhRvL.107k4504F} {107, 114504}

\bibitem[\protect\citeauthoryear{{Federrath}, {Sur}, {Schleicher}, {Banerjee}
  \& {Klessen}}{{Federrath} et~al.}{2011b}]{FSSBK11}
{Federrath} C.,  {Sur} S.,  {Schleicher} D. R.~G.,  {Banerjee} R.,   {Klessen}
  R.~S.,  2011b, \mn@doi [\apj] {10.1088/0004-637X/731/1/62}, \href
  {https://ui.adsabs.harvard.edu/abs/2011ApJ...731...62F} {731, 62}

\bibitem[\protect\citeauthoryear{{Fryxell} et~al.,}{{Fryxell}
  et~al.}{2000}]{Fry+00}
{Fryxell} B.,  et~al., 2000, \mn@doi [\apjs] {10.1086/317361}, \href
  {https://ui.adsabs.harvard.edu/abs/2000ApJS..131..273F} {131, 273}

\bibitem[\protect\citeauthoryear{{Gent}, {Mac Low}, {Korpi-Lagg}  \&
  {Singh}}{{Gent} et~al.}{2023}]{Gent+23}
{Gent} F.~A.,  {Mac Low} M.-M.,  {Korpi-Lagg} M.~J.,   {Singh} N.~K.,  2023,
  \mn@doi [\apj] {10.3847/1538-4357/acac20}, \href
  {https://ui.adsabs.harvard.edu/abs/2023ApJ...943..176G} {943, 176}

\bibitem[\protect\citeauthoryear{{Haugen}, {Brandenburg}  \& {Dobler}}{{Haugen}, 
{Brandenburg} \& {Dobler}}{2004}]{HBD04}
{Haugen} N.~E.,  {Brandenburg} A.,   {Dobler} W.,  2004, \mn@doi [\pre]
  {10.1103/PhysRevE.70.016308}, \href
  {https://ui.adsabs.harvard.edu/abs/2004PhRvE..70a6308H} {70, 016308}

\bibitem[\protect\citeauthoryear{{Kazantsev}}{{Kazantsev}}{1968}]{K68}
{Kazantsev} A.~P.,  1968, Soviet Journal of Experimental and Theoretical
  Physics, \href {https://ui.adsabs.harvard.edu/abs/1968JETP...26.1031K} {26,
  1031}

\bibitem[\protect\citeauthoryear{{Latif}, {Schleicher}, {Schmidt}  \&
  {Niemeyer}}{{Latif} et~al.}{2013}]{Latif+13}
{Latif} M.~A.,  {Schleicher} D.~R.~G.,  {Schmidt} W.,   {Niemeyer} J.,  2013,
  \mn@doi [\mnras] {10.1093/mnras/stt503}, \href
  {https://ui.adsabs.harvard.edu/abs/2013MNRAS.432..668L} {432, 668}

\bibitem[\protect\citeauthoryear{{Lee}}{{Lee}}{2013}]{Lee13}
{Lee} D.,  2013, \mn@doi [Journal of Computational Physics]
  {10.1016/j.jcp.2013.02.049}, \href
  {http://adsabs.harvard.edu/abs/2013JCoPh.243..269L} {243, 269}

\bibitem[\protect\citeauthoryear{{Lee} \& {Deane}}{{Lee} \&
  {Deane}}{2009}]{LD09}
{Lee} D.,  {Deane} A.~E.,  2009, \mn@doi [Journal of Computational Physics]
  {10.1016/j.jcp.2008.08.026}, \href
  {http://adsabs.harvard.edu/abs/2009JCoPh.228..952L} {228, 952}

\bibitem[\protect\citeauthoryear{{Marinacci} et~al.,}{{Marinacci}
  et~al.}{2018}]{Marinacci+18}
{Marinacci} F.,  et~al., 2018, \mn@doi [\mnras] {10.1093/mnras/sty2206}, \href
  {https://ui.adsabs.harvard.edu/abs/2018MNRAS.480.5113M} {480, 5113}

\bibitem[\protect\citeauthoryear{{Miyoshi} \& {Kusano}}{{Miyoshi} \&
  {Kusano}}{2005}]{MK05}
{Miyoshi} T.,  {Kusano} K.,  2005, \mn@doi [Journal of Computational Physics]
  {10.1016/j.jcp.2005.02.017}, \href
  {http://adsabs.harvard.edu/abs/2005JCoPh.208..315M} {208, 315}

\bibitem[\protect\citeauthoryear{{Moss} \& {Shukurov}}{{Moss} \&
  {Shukurov}}{1996}]{MS96}
{Moss} D.,  {Shukurov} A.,  1996, \mn@doi [\mnras] {10.1093/mnras/279.1.229},
  \href {https://ui.adsabs.harvard.edu/abs/1996MNRAS.279..229M} {279, 229}

\bibitem[\protect\citeauthoryear{{Pakmor} et~al.,}{{Pakmor}
  et~al.}{2017}]{Pak+17}
{Pakmor} R.,  et~al., 2017, \mn@doi [\mnras] {10.1093/mnras/stx1074}, \href
  {https://ui.adsabs.harvard.edu/abs/2017MNRAS.469.3185P} {469, 3185}

\bibitem[\protect\citeauthoryear{{Porter}, {Jones}  \& {Ryu}}{{Porter}, {Jones} 
 \& {Ryu}}{2015}]{PJR15}
{Porter} D.~H.,  {Jones} T.~W.,   {Ryu} D.,  2015, \mn@doi [\apj]
  {10.1088/0004-637X/810/2/93}, \href
  {http://cdsads.u-strasbg.fr/abs/2015ApJ...810...93P} {810, 93}

\bibitem[\protect\citeauthoryear{{Rincon}}{{Rincon}}{2019}]{R19}
{Rincon} F.,  2019, \mn@doi [Journal of Plasma Physics]
  {10.1017/S0022377819000539}, \href
  {https://ui.adsabs.harvard.edu/abs/2019JPlPh..85d2001R} {85, 205850401}

\bibitem[\protect\citeauthoryear{{Schekochihin}, {Maron}, {Cowley}  \&
  {McWilliams}}{{Schekochihin} et~al.}{2002}]{Sch+02}
{Schekochihin} A.~A.,  {Maron} J.~L.,  {Cowley} S.~C.,   {McWilliams} J.~C.,
  2002, \mn@doi [\apj] {10.1086/341814}, \href
  {https://ui.adsabs.harvard.edu/abs/2002ApJ...576..806S} {576, 806}

\bibitem[\protect\citeauthoryear{{Schekochihin}, {Cowley}, {Taylor}, {Maron}
  \& {McWilliams}}{{Schekochihin} et~al.}{2004}]{Scheko+04a}
{Schekochihin} A.~A.,  {Cowley} S.~C.,  {Taylor} S.~F.,  {Maron} J.~L.,
  {McWilliams} J.~C.,  2004, \mn@doi [\apj] {10.1086/422547}, \href
  {https://ui.adsabs.harvard.edu/abs/2004ApJ...612..276S} {612, 276}

\bibitem[\protect\citeauthoryear{{Schober}, {Schleicher}  \&
  {Klessen}}{{Schober}, {Schleicher} \& {Klessen}}{2013}]{SSK13}
{Schober} J.,  {Schleicher} D.~R.~G.,   {Klessen} R.~S.,  2013, \mn@doi [\aap]
  {10.1051/0004-6361/201322185}, \href
  {https://ui.adsabs.harvard.edu/abs/2013A&A...560A..87S} {560, A87}

\bibitem[\protect\citeauthoryear{{Seta} \& {Federrath}}{{Seta} \&
  {Federrath}}{2021a}]{SF21b}
{Seta} A.,  {Federrath} C.,  2021a, \mn@doi [Physical Review Fluids]
  {10.1103/PhysRevFluids.6.103701}, \href
  {https://ui.adsabs.harvard.edu/abs/2021PhRvF...6j3701S} {6, 103701}

\bibitem[\protect\citeauthoryear{{Seta} \& {Federrath}}{{Seta} \&
  {Federrath}}{2021b}]{SF21a}
{Seta} A.,  {Federrath} C.,  2021b, \mn@doi [\mnras] {10.1093/mnras/stab128},
  \href {https://ui.adsabs.harvard.edu/abs/2021MNRAS.502.2220S} {502, 2220}

\bibitem[\protect\citeauthoryear{{Seta}, {Bushby}, {Shukurov}  \&
  {Wood}}{{Seta} et~al.}{2020}]{Seta+20}
{Seta} A.,  {Bushby} P.~J.,  {Shukurov} A.,   {Wood} T.~S.,  2020, \mn@doi
  [Physical Review Fluids] {10.1103/PhysRevFluids.5.043702}, \href
  {https://ui.adsabs.harvard.edu/abs/2020PhRvF...5d3702S} {5, 043702}

\bibitem[\protect\citeauthoryear{{Seta}, {Rodrigues}, {Federrath}  \&
  {Hales}}{{Seta} et~al.}{2021}]{Seta+21}
{Seta} A.,  {Rodrigues} L. F.~S.,  {Federrath} C.,   {Hales} C.~A.,  2021,
  \mn@doi [\apj] {10.3847/1538-4357/abd2bb}, \href
  {https://ui.adsabs.harvard.edu/abs/2021ApJ...907....2S} {907, 2}

\bibitem[\protect\citeauthoryear{{Shukurov} \& {Subramanian}}{{Shukurov} \&
  {Subramanian}}{2021}]{SS21}
{Shukurov} A.~M.,  {Subramanian} K.,  2021, {Astrophysical Magnetic Fields:
  From Galaxies to the Early Universe}.
{Cambridge Univ. Press, Cambridge}

\bibitem[\protect\citeauthoryear{{Subramanian}, {Shukurov}  \&
  {Haugen}}{{Subramanian}, {Shukurov} \& {Haugen}}{2006}]{SSH06}
{Subramanian} K.,  {Shukurov} A.,   {Haugen} N.~E.~L.,  2006, \mn@doi [\mnras]
  {10.1111/j.1365-2966.2006.09918.x}, \href
  {http://adsabs.harvard.edu/abs/2006MNRAS.366.1437S} {366, 1437}

\bibitem[\protect\citeauthoryear{{Sur}}{{Sur}}{2019}]{Sur19}
{Sur} S.,  2019, \mn@doi [\mnras] {10.1093/mnras/stz1918}, \href
  {https://ui.adsabs.harvard.edu/abs/2019MNRAS.488.3439S} {488, 3439}

\bibitem[\protect\citeauthoryear{{Sur}, {Schleicher}, {Banerjee}, {Federrath}
  \& {Klessen}}{{Sur} et~al.}{2010}]{Sur+10}
{Sur} S.,  {Schleicher} D.~R.~G.,  {Banerjee} R.,  {Federrath} C.,   {Klessen}
  R.~S.,  2010, \mn@doi [\apjl] {10.1088/2041-8205/721/2/L134}, \href
  {https://ui.adsabs.harvard.edu/abs/2010ApJ...721L.134S} {721, L134}

\bibitem[\protect\citeauthoryear{{Sur}, {Federrath}, {Schleicher}, {Banerjee}
  \& {Klessen}}{{Sur} et~al.}{2012}]{Sur+12}
{Sur} S.,  {Federrath} C.,  {Schleicher} D. R.~G.,  {Banerjee} R.,   {Klessen}
  R.~S.,  2012, \mn@doi [\mnras] {10.1111/j.1365-2966.2012.21100.x}, \href
  {https://ui.adsabs.harvard.edu/abs/2012MNRAS.423.3148S} {423, 3148}

\bibitem[\protect\citeauthoryear{{Sur}, {Pan}  \& {Scannapieco}}{{Sur}, {Pan}
\& {Scannapieco}}{2014a}]{SPS14}
{Sur} S.,  {Pan} L.,   {Scannapieco} E.,  2014a, \mn@doi [\apj]
  {10.1088/0004-637X/784/2/94}, \href
  {https://ui.adsabs.harvard.edu/abs/2014ApJ...784...94S} {784, 94}

\bibitem[\protect\citeauthoryear{{Sur}, {Pan}  \& {Scannapieco}}{{Sur}, {Pan}
\& {Scannapieco}}{2014b}]{SPS14b}
{Sur} S.,  {Pan} L.,   {Scannapieco} E.,  2014b, \mn@doi [\apjl]
  {10.1088/2041-8205/790/1/L9}, \href
  {https://ui.adsabs.harvard.edu/abs/2014ApJ...790L...9S} {790, L9}

\bibitem[\protect\citeauthoryear{{Sur}, {Bhat}  \& {Subramanian}}{{Sur},
 {Bhat} \& {Subramanian}}{2018}]{SBS18}
{Sur} S.,  {Bhat} P.,   {Subramanian} K.,  2018, \mn@doi [\mnras]
  {10.1093/mnrasl/sly007}, \href
  {https://ui.adsabs.harvard.edu/abs/2018MNRAS.475L..72S} {475, L72}

\bibitem[\protect\citeauthoryear{{Sur}, {Basu}  \& {Subramanian}}{{Sur}, {Basu} \& 
{Subramanian}}{2021}]{SBS21}
{Sur} S.,  {Basu} A.,   {Subramanian} K.,  2021, \mn@doi [\mnras]
  {10.1093/mnras/staa3767}, \href
  {https://ui.adsabs.harvard.edu/abs/2021MNRAS.501.3332S} {501, 3332}

\bibitem[\protect\citeauthoryear{{Tobias}}{{Tobias}}{2021}]{T21}
{Tobias} S.~M.,  2021, \mn@doi [Journal of Fluid Mechanics]
  {10.1017/jfm.2020.1055}, \href
  {https://ui.adsabs.harvard.edu/abs/2021JFM...912P...1T} {912, P1}

\bibitem[\protect\citeauthoryear{{Vazza}, {Brunetti}, {Br{\"u}ggen}  \&
  {Bonafede}}{{Vazza} et~al.}{2018}]{Vazza+18}
{Vazza} F.,  {Brunetti} G.,  {Br{\"u}ggen} M.,   {Bonafede} A.,  2018, \mn@doi
  [\mnras] {10.1093/mnras/stx2830}, \href
  {https://ui.adsabs.harvard.edu/abs/2018MNRAS.474.1672V} {474, 1672}

\bibitem[\protect\citeauthoryear{{Xu} \& {Lazarian}}{{Xu} \&
  {Lazarian}}{2016}]{XL16}
{Xu} S.,  {Lazarian} A.,  2016, \mn@doi [\apj] {10.3847/1538-4357/833/2/215},
  \href {https://ui.adsabs.harvard.edu/abs/2016ApJ...833..215X} {833, 215}

\bibitem[\protect\citeauthoryear{{Xu} \& {Lazarian}}{{Xu} \&
  {Lazarian}}{2020}]{XL20}
{Xu} S.,  {Lazarian} A.,  2020, \mn@doi [\apj] {10.3847/1538-4357/aba7ba},
  \href {https://ui.adsabs.harvard.edu/abs/2020ApJ...899..115X} {899, 115}

\bibitem[\protect\citeauthoryear{{Yoon}, {Cho}  \& {Kim}}{{Yoon}, {Cho} 
\& {Kim}}{2016}]{YCK16}
{Yoon} H.,  {Cho} J.,   {Kim} J.,  2016, \mn@doi [\apj]
  {10.3847/0004-637X/831/1/85}, \href
  {https://ui.adsabs.harvard.edu/abs/2016ApJ...831...85Y} {831, 85}

\makeatother
\end{thebibliography}
\end{document}